\documentclass[12pt,preprint]{aastex}
\usepackage{emulateapj5,apjfonts}
\usepackage{graphics}
\usepackage{amssymb}
\usepackage{bbm}
\usepackage{amsmath,textcomp,array}
\usepackage{onecolfloat}
\usepackage[ruled]{algorithm}

\usepackage{bm}

\newcommand{\mb}[1]{\mathbf{#1}}

\newcommand{\Expect}[1]{\mathrm{E}\left[#1\right]}
\newcommand{\vol}{\text{vol}}
\newcommand{\abs}[1]{\left|#1\right|}
\newcommand{\norm}[1]{\left\|#1\right\|}

\newcommand{\nd}{d} 
\newcommand{\np}{n} 
\newcommand{\Design}{X} 
\newcommand{\region}{\mathcal{M}} 
\newcommand{\shift}{\mb p} 
\newcommand{\known}{\mathcal{K}} 
\newcommand{\Gen}{\mathbf G} 
\newcommand{\scale}{l} 

\newcommand{\RR}{\mathbb{R}} 
\newcommand{\ZZ}{\mathbb{Z}} 

\lefthead{Heitmann et al.}
\righthead{The Mira-Titan Universe: Precision Predictions for Dark
    Energy Surveys}

\begin{document}


\def\head{

  \title{The Mira-Titan Universe: Precision Predictions for Dark
    Energy Surveys} \author{Katrin~Heitmann\altaffilmark{1,2}, Derek
    Bingham\altaffilmark{3}, Earl Lawrence\altaffilmark{4}, Steven
    Bergner\altaffilmark{3}, Salman~Habib\altaffilmark{1,2}, David
    Higdon\altaffilmark{6}, Adrian Pope\altaffilmark{5}, Rahul
    Biswas\altaffilmark{1,7}, Hal Finkel\altaffilmark{5}, Nicholas
    Frontiere\altaffilmark{1,8}, Suman Bhattacharya\altaffilmark{1}}

\affil{$^1$ HEP Division,  Argonne National Laboratory, Lemont, IL
  60439, USA }
\affil{$^2$ MCS Division,  Argonne National Laboratory, Lemont, IL
  60439, USA } 
\affil{$^3$ Department of Statistics and Actuarial Science, Simon
  Fraser University, Burnaby, BC, Canada } 
\affil{$^4$ CCS-6, CCS Division, Los Alamos National Laboratory, Los
  Alamos, NM 87545, USA } 
\affil{$^5$ ALCF Division,  Argonne National Laboratory, Lemont, IL
  60439, USA }
\affil{$^6$ Social and Decision Analytics Laboratory, Virginia
  Bioinformatics Institute, Virginia Tech, Arlington, VA 22203, USA}   
\affil{$^7$ Department of Astronomy and the eScience Institute,
  University of Washington, Seattle, WA 98155, USA }    
\affil{$^8$ Department of Physics, University of Chicago, Chicago, IL
  60637, USA } 

\date{today}

\begin{abstract}
  Ground and space-based sky surveys enable powerful cosmological
  probes based on measurements of galaxy properties and the
  distribution of galaxies in the Universe. These probes include weak
  lensing, baryon acoustic oscillations, abundance of galaxy clusters,
  and redshift space distortions; they are essential to improving our
  knowledge of the nature of dark energy. On the theory and modeling
  front, large-scale simulations of cosmic structure formation play an
  important role in interpreting the observations and in the
  challenging task of extracting cosmological physics at the needed
  precision. These simulations must cover a parameter range beyond the
  standard six cosmological parameters and need to be run at high mass
  and force resolution. One key simulation-based task is the
  generation of accurate theoretical predictions for observables, via
  the method of emulation. Using a new sampling technique, we explore
  an 8-dimensional parameter space including massive neutrinos and a
  variable dark energy equation of state. We construct trial emulators
  using two surrogate models (the linear power spectrum and an
  approximate halo mass function). The new sampling method allows us
  to build precision emulators from just 26 cosmological models and to
  increase the emulator accuracy by adding new sets of simulations in
  a prescribed way. This allows emulator fidelity to be systematically
  improved as new observational data becomes available and higher
  accuracy is required. Finally, using one $\Lambda$CDM cosmology as
  an example, we study the demands imposed on a simulation campaign to
  achieve the required statistics and accuracy when building emulators
  for dark energy investigations.
\end{abstract}

\keywords{methods: statistical ---
          cosmology: large-scale structure of the universe}}

\twocolumn[\head]

\section{Introduction}

Cosmology has developed rapidly in the last few decades. The more
qualitative original picture of the evolution of the Universe has
evolved into a well-tested six-parameter ``Standard Model of
Cosmology''. Remarkably, this simple model describes all current
observations at a level of accuracy of roughly a few percent,
cementing a consistent picture of the Universe from a diverse set of
cosmological probes. While this progress has truly revolutionized our
knowledge of the Universe, a real understanding of its two main
constituents, dark matter and dark energy, remains elusive.

To make progress towards solving these mysteries, as well as shedding
light on other questions such as the sum of neutrino masses and the
nature of primordial fluctuations, new ground- and space-based
missions have been carried out (so-called Stage II experiments), are
ongoing (Stage III), and are under construction or planned (Stage
IV). These include the Baryon Oscillation Spectroscopic Survey
(BOSS)~\citep{boss}, WiggleZ~\citep{wigglez}, the Dark Energy Survey
(DES\footnote{https://www.darkenergysurvey.org/}), the Large Synoptic
Survey Telescope (LSST)~\citep{lsst}, the Dark Energy Spectroscopic
Instrument (DESI\footnote{http://desi.lbl.gov/}), the Wide Field
Infrared Survey Telescope (WFIRST)~\citep{wfirst}, and
Euclid~\citep{euclid}. The major science targets of these missions
specific to dark energy are measurements of the large scale structure
in the Universe on different scales (along with supernova distance
measurements). Weak lensing, baryon acoustic oscillations (BAO),
abundance of galaxy clusters, and redshift space distortions are the
most prominent probes aimed at investigating the nature of dark
energy. The upcoming surveys promise to obtain measurements of the
main cosmology parameters at the 1\% level accuracy if systematic
errors can be sufficiently controlled.

To reach the desired accuracies, the physics of matter clustering on
significantly nonlinear scales must be accurately modeled. Currently,
the only way to do this is via expensive numerical
simulations. Depending on the scales of interest and the required
accuracy, different physical effects must be accounted for. On large
scales relevant to, e.g., BAO, gravity-only simulations are likely
sufficient for interpreting ongoing and upcoming measurements. For
weak lensing, current surveys can be modeled with pure gravity
simulations (see, e.g., the discussion in \cite{coyote1}) but the
interpretation of future surveys will require a treatment of baryonic
effects (e.g., gas dynamics, feedback) and associated subgrid
modeling.

Besides adding hydrodynamics effects and other modeling ingredients to
the simulations, the cosmological parameter space has to be
simultaneously widened. As argued in \cite{coyote1}, current
observations do not have enough information to constrain cosmological
parameters beyond $w$CDM scenarios -- a mild extension of $\Lambda$CDM
-- at high accuracy. In the near future, however, it will be important
to go beyond $w$CDM by including, e.g., dynamical dark energy models
and at the same time accounting for ``secondary'' effects, currently
too small to reliably extract from the measurements beyond setting
upper or lower limits (such as a lower limit for $m_\nu$, the neutrino
mass sum). The work here is aimed at a research program that covers
the following eight-dimensional space:
$p_\theta=\{\omega_m,\omega_b,\sigma_8,h,n_s,w_0,w_a,\omega_\nu\}.$
(Details are found in Section~\ref{params}.)

\subsection{Simulation Campaigns}

If only a small number of simulations are required, the size of the
parameter space is essentially irrelevant. This is, however, not the
case in cosmology. Eventually one wishes to infer the values of the
parameters from a set of observational measurements. Commonly used
procedures for doing this all fall under the class of statistical
inverse problems, where a large number of evaluations of the forward
model (i.e., results from many simulations, at different specified
parameter values) are necessary in order to use Markov chain Monte
Carlo (MCMC) or other related methods. This may involve hundreds of
thousands or even millions of forward model evaluations. (A similar
issue is encountered in covariance estimation, where a large number of
simulations are also needed.)

At first sight this problem may appear to be computationally
overwhelming but it is in fact possible to cover large parameter
spaces in an efficient manner with a drastically reduced number of
simulations, provided certain smoothness conditions are met. The last
requirement is largely satisfied in cosmological inverse problems,
and this has allowed us to develop the ``Cosmic Calibration
Framework''
\citep{HHHN,HHHNW,coyote2,coyote3,higdon_oxford,emu_ext}. The major 
components of the framework are:

\begin{itemize}
\item Sophisticated sampling methods that provide optimal coverage of
  the parameter space. In previous work we used a symmetric Latin
  hypercube design, as explained in \cite{coyote2} (for a more
  technical presentation, see \citealt{santner03}). The new strategy
  presented here allows us to begin with a small set of cosmological
  models from which to construct a first set of emulators.  These
  simulations are then systematically augmented to improve emulation
  accuracy. This defines a notion of {\em strong} convergence
  (Cf. Section~\ref{design}) as opposed to the {\em weak} convergence
  previously demonstrated in \cite{HHHNW}. 
\item High-dimensional interpolater based on Gaussian process modeling
  to build a prediction scheme (the {\em emulator}) for a diverse set
  of cosmological statistics of interest, e.g., the fluctuation power
  spectrum, the halo mass function, lensing statistics, etc.
\item Emulator-based sensitivity analysis, characterizing the
  dependence of the cosmological statistics on different cosmological
  and modeling parameters.
\item Final calibration step, which combines observational data and
  theoretical predictions to extract cosmological information from the
  data via MCMC methods.
\end{itemize}
We will focus on the first three steps and construct precision
emulators for the power spectrum and the mass function; we also
discuss extensions to other already existing emulators, e.g., for the
concentration-mass relation~\citep{kwan13} and the galaxy power
spectrum and correlation function~\citep{kwan14} below.

In this paper, we narrow our focus to work that directly impacts
gravity-only simulations. The purpose here is to demonstrate the
success of the basic methodology. Over time, we expect to include
direct modeling of baryonic effects as well as indirect approaches
based on post-processing gravity-only simulations. As has been
mentioned previously, and discussed in several papers
\citep{white04,zhanknox,jing06,rudd,vanDaalen}, deriving reliable
predictions for cosmological statistics such as the power spectrum and
the mass function eventually requires the treatment of gas physics,
star formation, and feedback processes. As we explore smaller scales,
these effects become more important. For the matter power spectrum, as
an example, we expect effects at the 10-20\% level at scales of $k\sim
10~h$Mpc$^{-1}$. Accounting for these effects via simulations poses
two major difficulties: (i) the simulations are computationally very
expensive, so only a limited number can be carried out; (ii) the
physical models used in the simulations are too crude to be truly
predictive; there is a rather large uncertainty in the
modeling. Steady progress in this area is being made as hydrodynamical
simulations become more affordable and understanding of the influence
of baryonic physics improves.  Examples of recent attempts to mitigate
these uncertainties include the alteration of the halo
concentration-mass relation and exploring how this change feeds back
in, e.g., the power spectrum~\citep{zentner08}.  This approach can be
explored within the halo model to obtain predictions for the power
spectrum. More recently, it has been extended to include a range of
hydrodynamic simulations and the calibration of how baryonic physics
can bias the measurements from surveys like DES and
LSST~\citep{zentner13}. The idea is not to rely on carrying out a
hydrodynamic simulation that is accurate in all its details, but
rather to construct an approach that can mitigate uncertainties and
biases due to baryonic effects in robust ways.

Finally, a major challenge is posed by the need for accurate
predictions for covariances.  The main difficulty here is the very
large number of simulations that might be required to determine
covariances at the accuracy needed as well as the uncertainties due to
different cosmological inputs. Some early work in tackling these
questions via emulation has been carried out
by~\cite{SKHHHN}. \cite{morrison13} focus in particular on the
question of cosmology dependence and emulation, basing their analysis
on a halo model approach. As for baryonic effects, more work is
required but first results are encouraging. We will not address either
of these problems here but stress that our general emulation approach
is directly applicable to these problems.

To attack the aforementioned challenges a multi-step comprehensive
program has to be designed and carried out: (i) calibrate the
underlying dark matter power spectrum and mass function for an
8-parameter model space out to $k\sim 5$Mpc$^{-1}$ and mass between
$2\cdot 10^{12}$M$_\odot$ and $10^{15}$M$_\odot$ (at $z=0$, smaller
cut-offs for high masses at higher redshifts) at high accuracy; (ii)
extend the $k$-range to higher $k$ and mass range to lower masses by
augmenting the simulation suite with a small number of simulations
with very high mass resolution and use smart extrapolation approaches
where applicable; (iii) add a small number of high-fidelity
simulations including gas physics and re-calibrate the power spectra
and mass functions for these effects; (iv) gauge the remaining
uncertainties in the predictions and fold them into the calibration
step.

Such a program is obviously very ambitious and will require
significant effort and resources to be completed. In addition,
community support and input are crucial -- the simulation suite that
will result from such a project can be used for a large number of
scientific projects and has to be made available to the entire
community. In order to accommodate as many follow-up projects as
possible, it is important to have most analysis tools already in place
and a comprehensive plan for post-processing so that the relevant data
can be properly archived.

The goal of the current paper is to focus on the first step in some
detail and demonstrate with surrogate models (such as the linear power
spectrum and an estimated mass function) and a set of high resolution
simulations that it is possible to successfully complete this program
in accordance with the timelines of the major surveys. The paper will
provide a roadmap that delivers prediction schemes at different levels
of accuracy at different times -- culminating finally in high-accuracy
emulators that will be required for extracting the most science from
surveys such as LSST. It introduces a new statistical approach for
designing a simulation campaign that will lead to improved emulators
over time. We also show results from simulations to investigate the
optimal size of the simulations to be carried out, considering limited
computational resources as well as accuracy requirements. We strongly
believe that the community has to discuss and add to this roadmap to
make such a project successful. Therefore, a paper outlining the plan
and the solutions to the problem and showing that it is feasible
before embarking on this endeavor is crucial.

The paper is organized as follows. In Section~\ref{params}, we discuss
the parameter ranges covered by our study and how we derived them. We
then introduce a new design strategy for a simulation campaign that
can be augmented over time and will lead to higher accuracy emulators
with each new set of simulations (Section~\ref{design}).  A short
overview on the N-body simulations that will be employed for the
project is presented in Section~\ref{nbody}.  In the following
sections, we focus on precision predictions for the dark matter power
spectrum (Section~\ref{power}) and the halo mass
function~(Section~\ref{massf}) using this new strategy.  We present
convergence studies and demonstrate that a set of $\sim$100
cosmological models suffice to reach the required accuracy. We also
use a set of high resolution simulations to investigate the
requirements for each individual model to mitigate uncertainties due
to realization scatter. In Section~\ref{beyond} we summarize
additional emulator projects that can be envisioned and conclude in
Section~\ref{conclusion} with a summary and short discussion.

\section{Parameter Ranges}
\label{params}

As discussed in the Introduction, a major challenge posed by upcoming
surveys is the increase in parameter space that must be
handled. In~\cite{coyote2}, we discussed the requirements for ongoing
weak lensing surveys and concluded that a five dimensional parameter
space with $p_\theta=\{\omega_m,\omega_b,\sigma_8,n_s,w_0\}$ was
sufficient to capture the information from contemporary
measurements. In this scenario, we assume flatness, no running of the
spectral index $n_s$, massless neutrinos, and constraints on the
Hubble parameter $h$ from measurements of the distance to last
scattering.

In this paper, we build upon the results obtained in~\cite{coyote2}
and Planck constraints~\citep{planck} and increase the parameter space
by three additional parameters.  We extend our previous analysis by
letting $h$ be a free parameter, allow for a dynamical dark energy
component via a two-dimensional parameterization $(w_0,w_a)$, and
allow for massive neutrinos via $\omega_{\nu}$ but still keep the
assumption of three relativistic species. For the first five
parameters we keep the same parameter ranges as in \cite{coyote2} and
we refer the reader to that paper for more details on the reasoning
behind our choices. We provide justifications for the chosen
additional parameter ranges below.  With the Planck data at hand, we
keep the assumptions of flatness, $\Omega_k=0$, and no running of the
spectral index, $d\log n_s/d\log k=0$. In summary, we investigate the
following eight parameters within the ranges:

\begin{figure}[t]
\centerline{
 \includegraphics[width=2.5in]{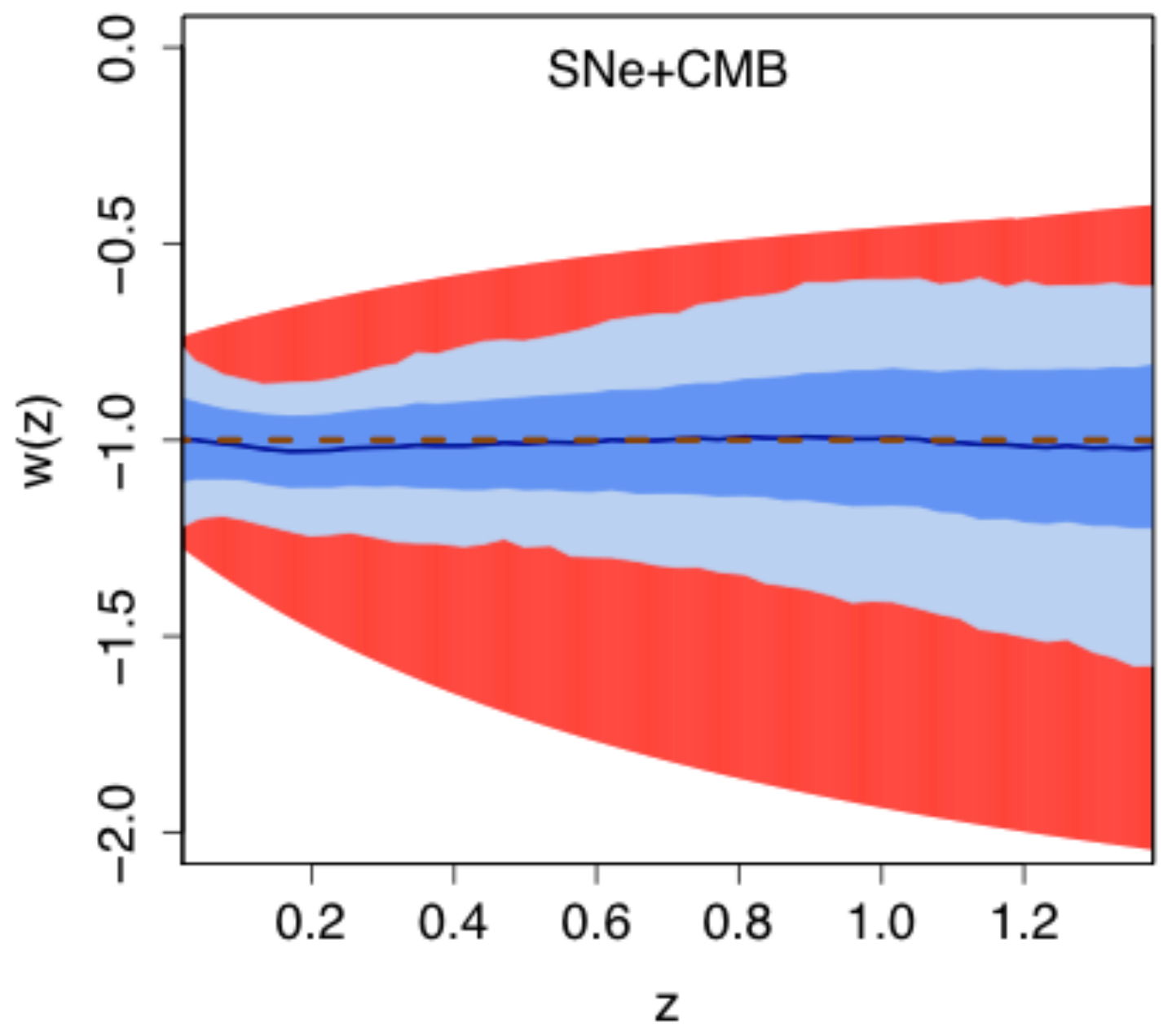}}
\caption{\label{wzband}Constraints for the dark energy equation of
  state parameter $w(z)$ from supernova and CMB data in blue (light
  blue: 95\% confidence level, dark blue: 68\% confidence level, red
  dashed line: $w=-1$) from \cite{holsclaw11}. The inclusion of baryon
  acoustic oscillations would tighten the constraints further. The red
  shaded region shows the range for $w(z)$ covered by our choices for
  $w_0$ and $w_a$ for our main design, comfortably including the best
  current constraints. }
\end{figure}

\begin{eqnarray}\label{cosmoparams}
0.12\le &\omega_m& \le 0.155\\
0.0215\le &\omega_b& \le 0.0235\\
0.7\le &\sigma_8& \le 0.9\\
0.55\le &h& \le 0.85\\
0.85\le &n_s& \le 1.05\\
-1.3\le &w_0& \le -0.7\\
-1.5\le &w_a& \le 1.15\\
0.0\le &\omega_\nu& \le 0.01.
\end{eqnarray}

For these parameter ranges our aim is to obtain high precision
predictions for different observables, at the percent level of
accuracy.

\begin{figure}
\centerline{
\includegraphics[width=3.3in]{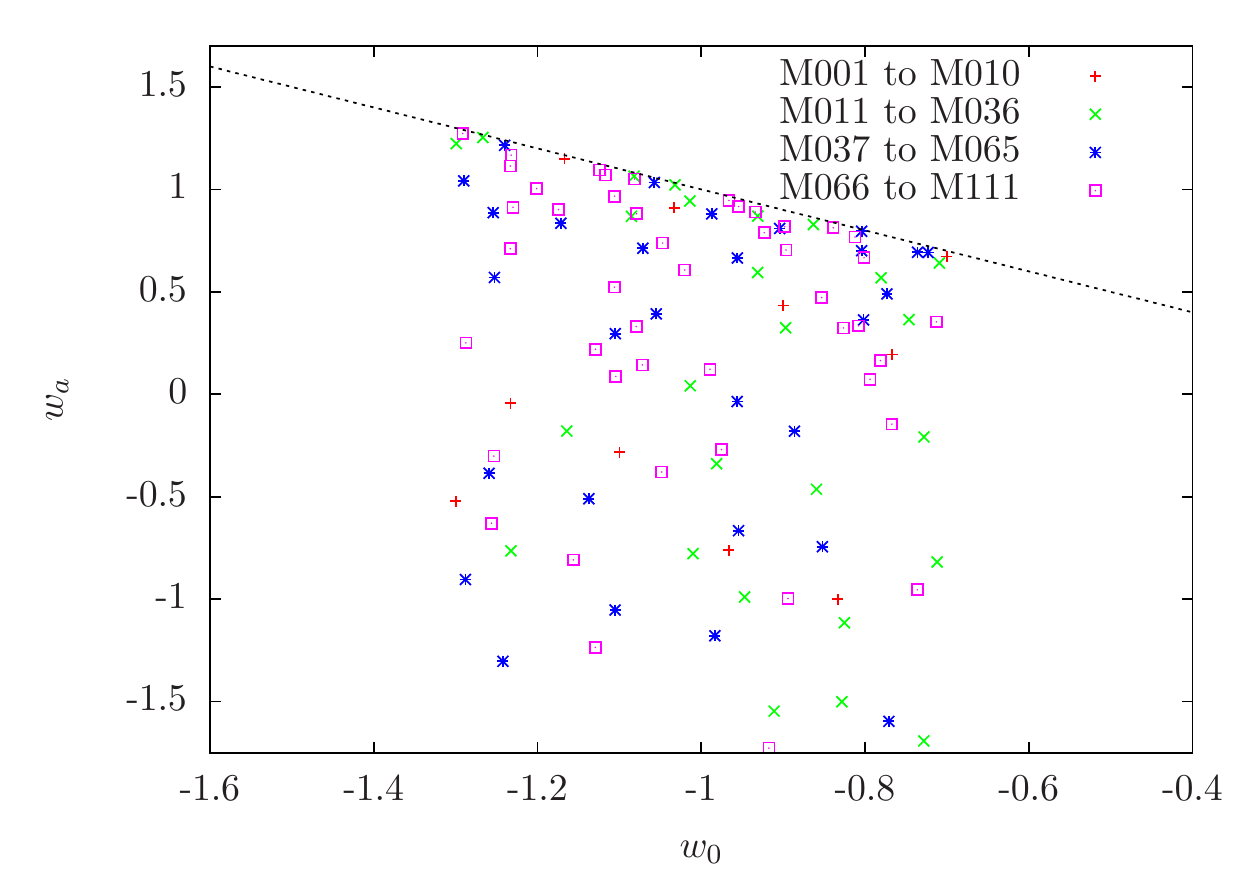}}
\caption{\label{w0wa}Coverage of the $(w_0,w_a)$ plane. Red crosses
  show the design points with zero neutrino masses, green crosses the
  first 26 models, blue stars the next set of 29 models, and pink
  boxes the remaining 46 models. The upper dashed line shows the limit
  $w_0+w_a < 0$ obeyed by all the models. The area shown approximately
  covers the constraints obtained from Planck data combined with the
  Union2.1 or SNLS supernova data set (see Fig.~36 in
  \citealt{planck}).  Models above the dotted line ($w_0+w_a = 0$) are
  not considered. }
\end{figure}

\begin{figure}[t]
\centerline{
 \includegraphics[width=3.3in]{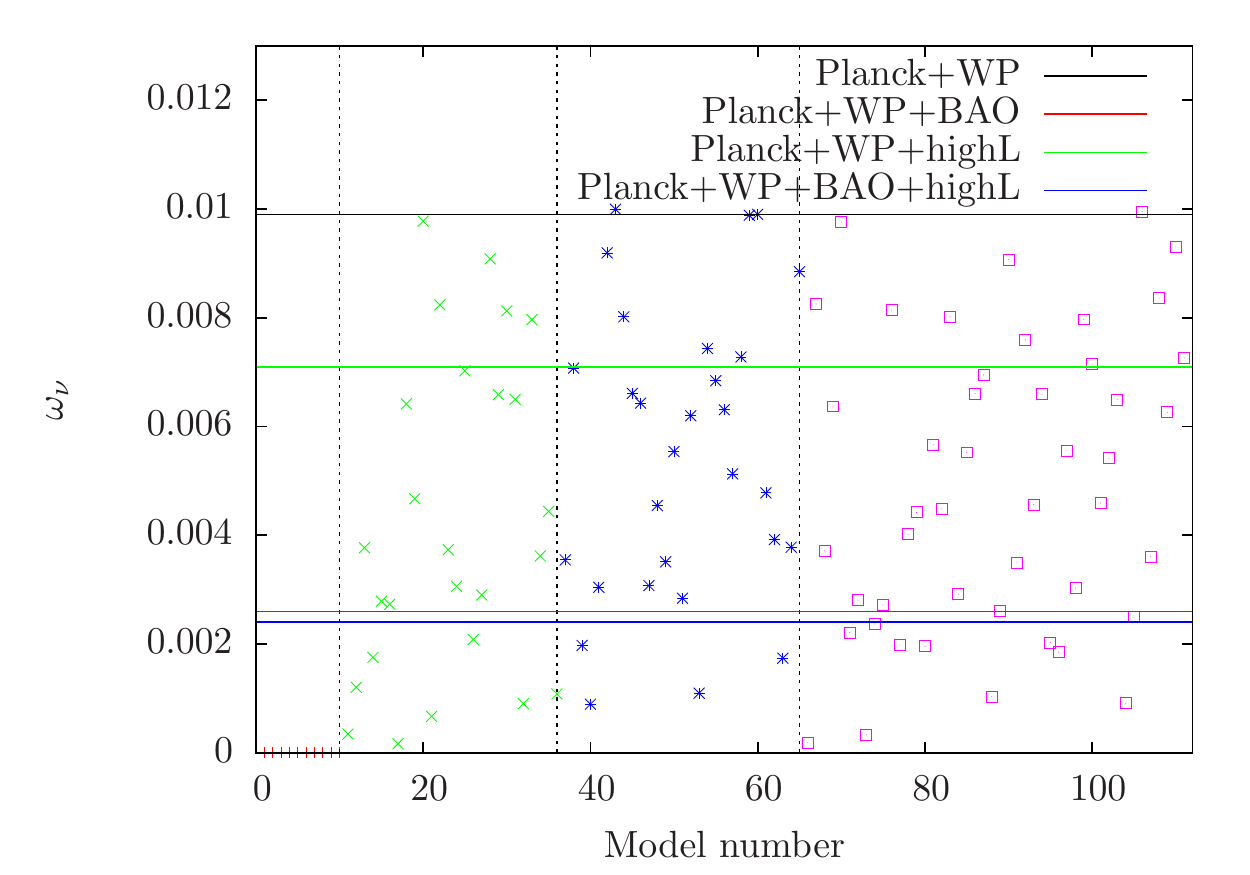}}
\caption{\label{neutr}Current upper bounds on $\omega_\nu$ from Planck
  and combinations of different cosmological probes for a $\Lambda$CDM
  cosmology (values from Table~10 in \citealt{planck}). WP stands for
  WMAP polarization data, the BAO measurements include several
  different surveys, the `highL' CMB measurements are from the Atacama
  Cosmology Telescope and the South Pole Telescope.  The red symbols
  show the values for $\omega_\nu$ in our design for the 101+10 models
  we consider throughout the paper. The first ten models have
  $\omega_\nu=0$ to provide good coverage of the current Standard
  Model. The symbols have the same meaning as in Fig.~\ref{w0wa}. }
\end{figure}

In \cite{coyote2}, we determined the best fit value for the Hubble
parameter $h$ for each model from CMB measurements of the distance to
last scattering. The power spectrum emulator \citep{coyote3} simply
output the best fit value for any other emulated model. In the
extended Coyote emulator (\citealt{emu_ext}) we provide the power
spectrum prediction out to higher $k$- and $z$-values as well as treat
the Hubble parameter as a free parameter in an approximate way. This
extended emulator is based on the same cosmological models as the
original emulator, which did not allow us to provide sub-percent
accurate predictions including $h$ as a free parameter.  Instead, we
used results from higher order perturbation theory to follow the
behavior of the power spectrum under changing $h$ (for sufficiently
small values of $k$).

In the current paper, we keep $h$ as a free parameter embedded in the
full design.  We choose the range according to the minimum and maximum
value we found in the model space in \cite{coyote2}. This range is
rather large with respect to currently available constraints on $h$ --
recent measurements shown in \cite{riess11} constrain $H_0$ at the 3\%
accuracy level to $73.8\pm2.4$~km/s/Mpc, suggesting that measurements
at the level of 1\% accuracy are possible -- but we want to ensure
that the previously investigated parameter space is comfortably
embedded in the new one.  In addition, the rather low value for
$H_0=67.3\pm1.2$~km/s/Mpc found by the Planck team suggests that the
error estimate in \cite{riess11} might be underestimated.

Next, we discuss our choices for the range of $w_0$ and $w_a$. The
uncertainty in these parameters, in particular $w_a$, is still very
large and reducing it is a major target of the upcoming surveys. We
therefore allow a large range for both parameters, informed by
constraints from supernova and CMB data.  In Fig.~\ref{wzband} we show
constraints obtained from a non-parametric reconstruction approach for
$w(z)$ using recent supernova and CMB constraints
(see~\citealt{holsclaw11} for details).  The blue regions show the
constraints at the 95\% and 68\% confidence level while red shaded
region shows the coverage by our main models.  Current constraints are
therefore comfortably covered by our allowed parameter ranges. A
similar view, but this time aimed at current parametric constraints on
$(w_0,w_a)$ is shown in Fig.~\ref{w0wa}. Here we show the coverage of
the $(w_0,w_a)$ parameter plane and scale the axis approximately
following Fig.~36 in \cite{planck} which provides constraints from CMB
and supernova measurements. Again, with the ranges as chosen, we
provide good coverage of the currently favored regions of parameter
space.

Finally, our choice for the range on the neutrino mass parameter
$\omega_\nu$ is informed by the constraints available from the Planck
measurements. In Fig.~\ref{neutr} we show different limits given by
Planck in combination with other surveys. Note that these limits were
derived for $\Lambda$CDM models. For our broader model space they
would be less stringent. We therefore choose to follow the most
conservative bound (Planck + WP). The WMAP team has considered
neutrino mass bounds for $w$CDM models and even for those our chosen
parameter range comfortably covers all currently allowed mass values.

\section{Design Strategy}
\label{design}

An important challenge for building efficient emulators in the
nonlinear regime is overcoming the high cost of accurate
simulations. It is not possible to evaluate the simulator over and
over in a brute force fashion. Given this hard constraint, a judicious
choice of simulation design (i.e., the simulation suite of models) is
essential.  For us, this means varying the eight parameters over the
ranges specified in Eqns.~\ref{cosmoparams} so that the parameter
space is optimally sampled.

The emulation strategy relies heavily on Gaussian process (GP) models
(see, e.g., \citealt{sacks89}; \citealt{Rasmussen06}) to interpolate
across the simulation results.  \cite{Johnson:1990:minimax} showed
that under certain conditions, designs with good space-filling
properties (e.g., maximizing the minimum distance between the design
points - maximin designs) are optimal for GPs. This is because a
prediction from a GP emulator is a weighted average of the data inputs
(simulation inputs in our case), where the outputs from nearby models
have higher weights than those far from the prediction location.

A very important new feature with our simulation design is that we aim
to progressively fill the model space so that intermediate analyses
can be performed while the remaining simulations continue to run.
Hence, a simulation design is required that simultaneously combines
(i) good space-filling properties for the chosen number of sampling
points and (ii) also provides good space-filling properties at
intermediate stages of the sampling process.  Together (i) and (ii)
ensure that the final and intermediate stage simulation designs give
good coverage of the model space so that efficient emulators can be
constructed in a staged, convergent fashion.  To achieve this goal, we
use a space-filling design based on point lattices. 

It is worth noting that Latin hypercube designs~\citep{mckay79} are
frequently used in simulation based ``experiments'', as in our
previous work.  These designs impose one-dimensional space-filling.
To make them better suited to computer model emulation with GPs,
additional space-filling criteria, e.g., maximin Latin hypercube
designs~(\citealt{morris95}), are imposed to improve the design.  The
nested lattice scheme we use here has the added benefit of explicitly
enforcing space-filling properties in intermediate stages of the
simulation campaign as well as in the final design.
 
\subsection{Space-filling lattice designs} 

A \emph{point lattice} $\Lambda$ is an infinite set of points in
$\RR^\nd$ constructed as integer multiples of a set of basis vectors
that are given as columns of a non-singular \emph{generating matrix}
$\;\Gen\in \RR^{\nd \times \nd}$
\begin{equation}
\Lambda(\Gen)=\{\Gen\mathbf{k} : \mathbf{k} \in \mathbb{Z}^\nd\}
\subset \RR^\nd, 
\label{eq:lattice}
\end{equation}
writing $\Lambda(\Gen)$ instead of $\Lambda$ only when the generating
matrix is needed to differentiate between lattices.  The familiar
Cartesian lattice, for example, has the $d$-dimensional identity
matrix as its generating matrix.

Since a lattice $\Lambda$ is a linear transformation of the integers
$\mathbb{Z}^\nd$, $\Lambda$ inherits the integer's Abelian group
structure, i.e., for any point $\mathbf{x} \in \Lambda$ we have
$\mathbf{x} + \Lambda = \Lambda$.  A \emph{Voronoi cell} is the
neighbourhood of points in $\RR^\nd$ that are closer to a particular
point $\mathbf{x}\in\Lambda$ than to any other point in $\Lambda$.
Due to the translation invariance of $\Lambda$, its Voronoi cells look
exactly the same for any $\mb x \in \Lambda$. Hence, a lattice basis
$\Gen$ that is chosen to give a large packing sphere diameter among
nearest neighbours (maximin distance) or small covering diameter
(minimax distance) immediately imposes useful space-filling properties
for the entire design region $\region$ as illustrated in Figure
\ref{f:designs} (\citealt{ConwaySloane:1999,Johnson:1990:minimax}).
It is traditional to scale all input parameters to the unit interval
$[0,1]$ to put all parameters on the same footing in the analysis.
Thus, it is convenient to write the input region as the
$\nd$-dimensional unit cube $\region = [0,1]^\nd$ with volume $\vol
(\region) = 1$.

A \emph{lattice design} is the intersection of the lattice and the
design region (or input region) of interest, possibly shifted to
achieve useful statistical properties.  More formally,
$\Design=\region \cap \Lambda + \mb p$ for input region $\region$ is
obtained by restricting the infinite lattice to the region with a
possible shift of $\Lambda$ by some $\mb p \in \RR^\nd$. A
straightforward approach for finding a lattice design is given in
Algorithm \ref{alg}.
\begin{algorithm}
  \caption{\label{alg}A method to produce a \emph{lattice design}
    $\Design(\region,\Gen,\shift)$ in the region $\region=[0,1]^\nd$
    for any generating matrix $\Gen$ and shift of origin
    $\shift$. \label{alg:makeDesign}}
\begin{enumerate}
\item Determine a finite candidate subset of lattice point indices $K'
  \subset \ZZ^\nd$ that produce all design points that are potentially
  contained in $[0,1]^\nd$. \\ 
  {\footnotesize For instance, enumerate all integer points in the
    axis-aligned bounding box that contains the inverse image $C'$ of
    all corners of the design region $\region$,
    $C'=\left\{\Gen^{-1}(\mb c - \shift) : \mb c \in
      \{0,1\}^\nd\right\}$.  With lower bound $\mb l \in \ZZ^\nd$ that
    is constructed as $l_i = \lfloor \min \{ k_i : \mb k \in C'
    \}\rfloor$ and analogous upper bound $\mb u \in \ZZ^\nd$, the
    candidate indices are $K' = \{l_1,l_1+1,\dotsc,u_1\} \times
    \{l_2,l_2+1,\dotsc,u_2\} \times \dots \times
    \{l_\nd,l_\nd+1,\dotsc,u_\nd\}$.  }
  \item Produce a superset of the design as $\Design' = \left\{
      \Gen\mb k + \shift : \mb k \in K' \right\}$. 
  \item Keep all points from $\Design'$ with coordinates $0\le x_{i,j}
    \le 1$ to obtain the design $\Design$. 
  \end{enumerate} {\footnotesize Note, to be able to produce lattice
    designs for larger $\nd$ it is essential to improve the
    construction of $K'$ in step 1. To sketch the approach here, start
    by covering a one-dimensional projection of
    $\Gen^{-1}(\region-\shift)$ to build the design iteratively,
    adding one dimension at a time.  }
\end{algorithm} 

\subsection{Searching for a good lattice design}
The choice of $\Gen$ and $\shift$ in Algorithm~\ref{alg:makeDesign}
leaves only $\nd^2+\nd$ degrees of freedom in optimizing further
useful experimental design criteria. This is important to finding good
designs for large dimensionality $\nd$ and simulation experiment
\emph{runsizes} $\np=\abs{\Design}$, because without the lattice
constraint this would be a $\nd\cdot\np$-dimensional, non-convex
optimization problem.  Furthermore, known constructions for good
generating matrices $\Gen$ are readily
available~(\citealt{ConwaySloane:1999}), leaving only a suitable
rotation $\mb Q \in \mathrm{SO}(\nd)$ to produce alternative bases
$\mb Q\Gen$.

When a lattice is shifted or rotated, design points can move in and
out of the fixed design region $\region$.  The \emph{expected
  runsize}, $\hat{\np}$, of a randomly rotated and shifted lattice
within the unit cube $\region$ is $\hat{\np}(\Gen) =
\Expect{\abs{\Design(\region, \mb Q\Gen,\shift)} \;:\; \shift \in
  \RR^\nd, \mb Q^T\mb Q = \mb I} = 1 / \abs{\det \Gen}$.  To achieve a 
certain desired runsize $N$, at least in expectation, we scale $\Gen$
such that $\det \Gen = 1/N$, before performing
Algorithm~\ref{alg:findDesign}.
\begin{algorithm}[t]
\caption{A method to find a shift $\shift$ and initial rotation $\mb
  Q_0$ of a lattice $\Lambda$ that produce a design with a desired
  runsize $N=1/\det \Gen$ and other possible design
  criteria.\label{alg:findDesign}} 
\begin{enumerate}
  \item Start with an empty set of known designs $\known$.
  \item Do the following for a certain number of trials (batch size). 
	\begin{enumerate}
	  \item Choose a random shift $\shift \in \region$ and produce
            a random rotation $\mb Q$ via QR-decomposition of a random
            matrix. 
	  \item Generate the lattice design $\Design(\region,\mb
            Q\Gen,\shift)$ via Algorithm~\ref{alg:makeDesign}. 
	  \item Calculate all design performance criteria
            $c_\scale(\Design)$ and store them as \emph{assessment}
            $\mb c$. This may just include the desired runsize $N$
            criterion $c_\text{runsize} = \abs{(\abs{\Design}-N)}$. 
	  If $\norm{\mb c}=0$, terminate the search and return design
          $\Design$ with rotation $\mb Q_0$ and $\shift$. 
	  \label{def:critl}
	  \item Add the tuple $(\mb Q, \shift, \mb c)$ to the current
            batch of trial designs. 
	\end{enumerate}
  \item Add the current batch of designs to the set of known designs
    $\known$.\\ 
  Repeat step 2, if the assessments $\mb c$ of known designs $\known$
  have changed significantly, for instance, as determined by the
  $\mathcal{L}_1$ distance of the percentile functions of empirical
  distributions of $\norm{\mb c}$ in $\known$ vs. $\known$ joined with
  the current update batch of trials.    
  \item Return one of the best known designs, i. e. by choosing one
    with the smallest $\norm{\mb c}$. 
\end{enumerate}
\end{algorithm}

\subsection{Sequences of nested lattice designs}

A further desirable property of a sequence of design points is that
they approach a uniform distribution --- ideally ensuring good
coverage for prediction at earlier run sizes.  This means, for
example, that even with a potential simulation budget of $200$
simulation runs, the first $25$ could be chosen so that they can
inform a decision as to whether resources should be allocated to
compute a further $25$ runs, then $50$, and then another $100$
experimental points to gain further insight.

In this section, we show how to obtain a sequence of lattice designs
that progressively fills the parameter (input) space.  First, observe
that a sequence of nested lattices is a hierarchy of lattices.  To
decompose a fine lattice $\Lambda(\Gen)$ into coarser lattices, the
generating matrix changes into $\Gen\mb K$ for {\em dilation matrix}
$\mb K$.  Central to the construction of {\em{nested}} lattice designs
is that $\Lambda(\mb G_0)\supset\Lambda(\mb G_1) \Leftrightarrow \mb
G_0^{-1} \mb G_1 = \mb K \in \ZZ^{\nd\times\nd}$.  In other
words, any sequence of nested lattices 
\begin{equation} 
\dotsc\supset\Lambda(\Gen_0)\supset\Lambda(\Gen_1)\supset\dotsc 
\label{eq:nesting}   
\end{equation}
is equivalent to a sequence of dilation matrices $\mb K_\scale$
{\em{being integer}}, where $\Gen_{\scale+1} = \Gen_\scale\mb
K_\scale$.  The initial lattice $\Lambda(\Gen_0)$ itself, however,
does not have to be a subset of the integers.

The expected runsize $\hat{\np}$ or {\em{density of $\Lambda(\Gen)$}}
is increased by a factor of $\abs{\det \mb K}=\beta\in\ZZ$ as opposed
to the coarser sub-lattice $\Lambda(\Gen\mb K)$.  It is possible to
choose the factor $\beta$ as low as two.  So, we can construct a
sequence of lattice designs with expected run-sizes that change by a
factor of two.

Here we provide a new method for constructing a sequence of nested
lattice designs.  This is achieved by developing a construction for a
single dilation matrix, $\mb K$, so that $\mb K^{-1}$ can be used many
times to obtain refined designs that yield large run sizes $\np$. That
is, we start with a coarse lattice, and apply the inverted dilation
matrix, to get larger designs.  The larger lattice designs contain the
smaller lattice designs and thus have good space-filling properties at
intermediate run sizes if performed in a sequence.

More precisely, our construction is giving all non-equivalent integer
matrices $\mb K$ that give a sequence of lattices
\begin{equation}
\dotsc\supset\Lambda(\Gen\mb
K^{-1})\supset\Lambda(\Gen)\supset\Lambda(\Gen\mb K)\supset\dotsc
\nonumber 
\end{equation}
that also have a rotational symmetry
\begin{equation}
\mb Q \Gen = \mb \Gen \mb K
\label{eq:rotational-nesting}
\end{equation} 
for a uniformly $\alpha$-scaled rotation matrix $\mb Q$ that fulfills  
$\mb Q^T\mb Q=\alpha^2\mb I$ with $\alpha^\nd=\det \mb Q=\det \mb K    
$.

\begin{figure}[t]
\centering
\includegraphics[scale=0.5, trim=10 5 5
10]{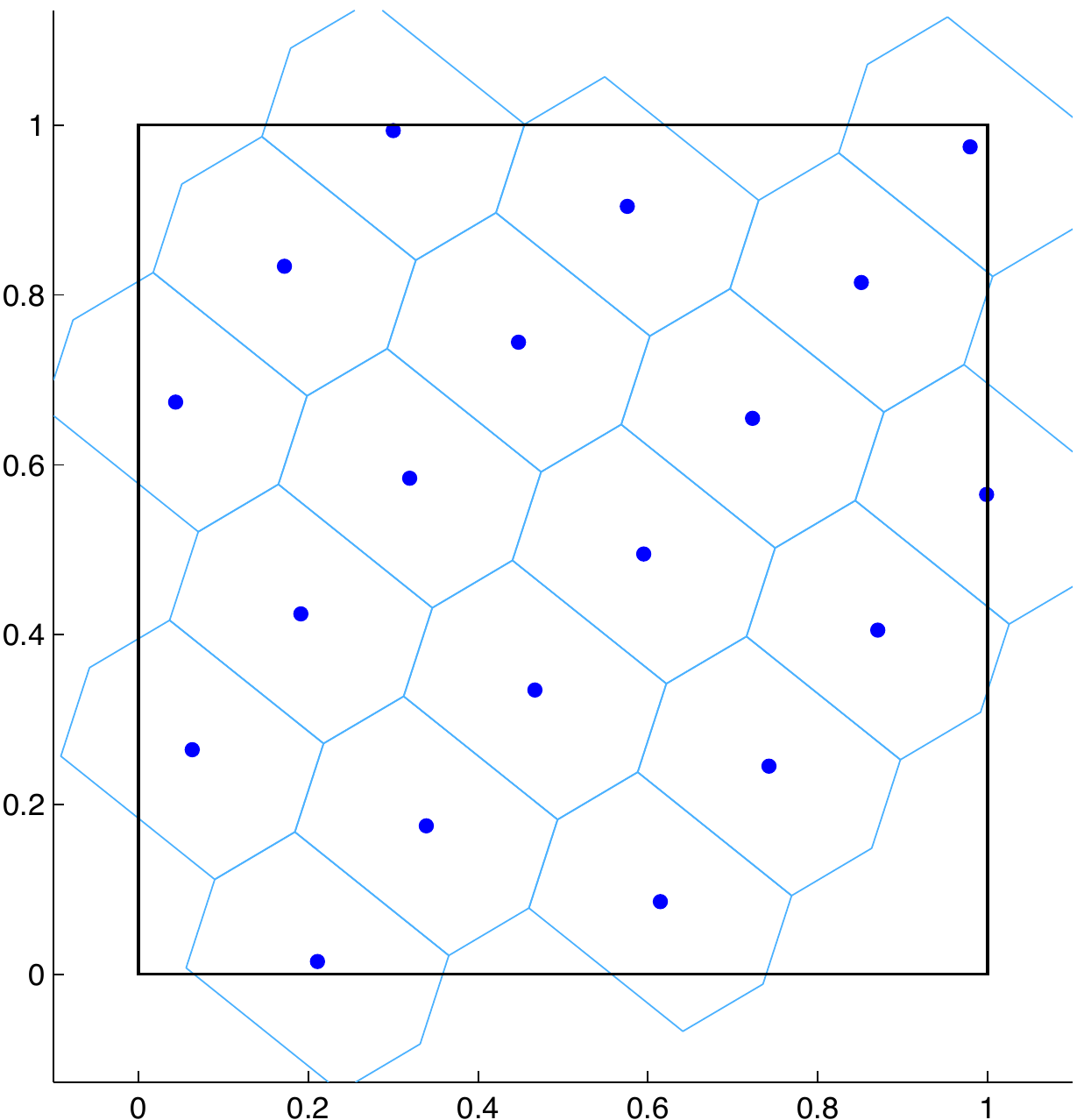} 
\hspace{0.5cm}
\includegraphics[scale=0.5, trim=10 5 5
10]{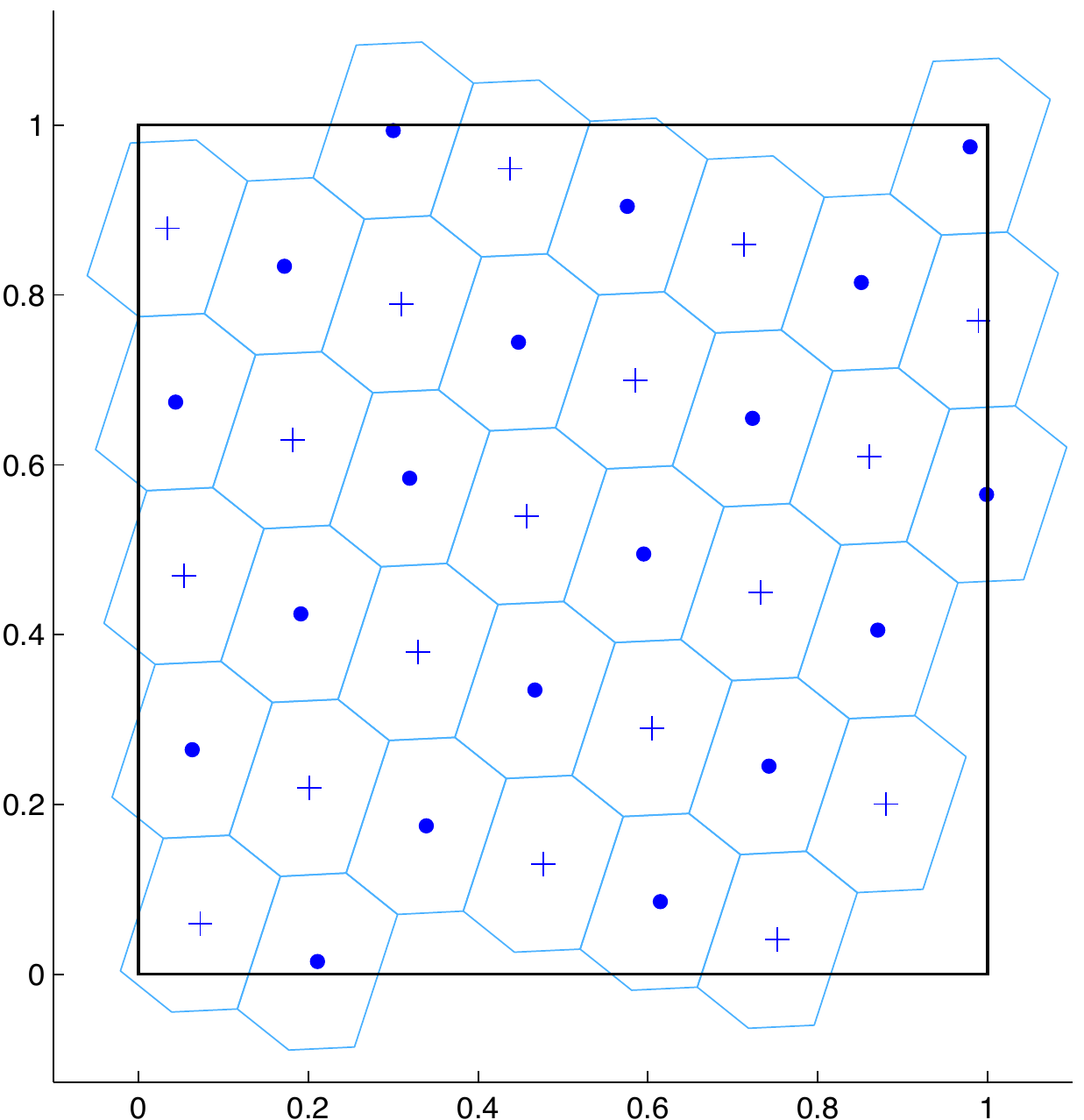} 
\caption{The 19 point set (a) is a subset of the 37 point refined
  design (b), which is a rotated version of (a) that has twice the
  density of sample points. 
}
\label{f:designs}
\end{figure}

Equation~\ref{eq:rotational-nesting} suggests an equivalent viewpoint
to our construction --- the sequence of \emph{nested} designs is
constructed from \emph{uniformly scaled rotations} $\mb Q$ of a
specific initial lattice $\Lambda(\mb G)$.  To find a suitable
initialization $\Design$ for the design sequence, the generating
matrix $\Gen$ can be scaled and rotated to optimize some criteria
using Algorithm~\ref{alg:findDesign}.  Coarser sub-designs
$\Design_\scale$ based on $ \Lambda(\mb Q^\scale \mb G)$ or
equivalently $\Lambda(\Gen\mb K^\scale)$ for $\scale=1,2,\dotsc$, as
well as refined designs for negative $\scale$ can be obtained via
Algorithm~\ref{alg:makeDesign}.

A hierarchical ordering of points is obtained by decomposing $\Design$
and by running the points in $\Design_\scale$ without repetition in
decreasing order starting with the largest $\scale$.  In practice, the
\emph{$\scale^\text{th}$ sub-grid} $\Design_{\scale}$ is determined by
retaining all points of $\{\mb K^{-\scale}\mb k: \mb k \in
\Gen^{-1}(\mb x-\shift): \mb x \in \Design\}$ that have integer
coordinates only.  A one-dimensional example $\Design_1$ with $\mb
K=2$ are the \emph{even numbers} formed from integer numbers that
after a multiplication by $\mb K^{-1}$ are still integer.

The choice of generating matrix $\Gen$ and rotation-scale matrix
$\mathbf{Q}$ (Eqn.~\ref{eq:rotational-nesting}) to produce a nested
design sequence are not independent.  \cite{Bergner:2011} demonstrated
that, given $d$ and sub-sampling rate $\beta=\det \mb K$, there are a
limited number of integer matrices $\mb K$ with corresponding
rotation-scale matrices, $\mathbf{Q}$, and $\Gen$, that can be
considered --- only four possibilities, $\mb K$ and $\mb K^T$ with
possible scaling by $-1$, exist in case of odd $\nd$ and prime
$\beta$.

The lattices obtained in this manner have good space-filling
properties that lead to efficient prediction for GPs.  In addition,
the sequences from one coarse lattice to increasingly refined lattices
of the \emph{same type} gives very predictable improvement for the
space filling properties (see illustration below).

Furthermore, unlike the Cartesian lattice where the size of the
simulation suite grows exponentially with $\nd$, the non-rectangular
lattices we use can achieve virtually any run-size.  Perfect ratios
$\beta$ among runsizes $|\Design_\scale|$ are not guaranteed, because
some of the points in the refined design can lie outside of $\region$.
With Algorithm~\ref{alg:findDesign} it is possible in
step~(\ref{def:critl}) to simultaneously match runsizes on multiple
levels of coarsening $\scale$, as additional criteria $c_\scale$.

As an illustration, consider the 2-d example in
Figure~\ref{f:designs}. Optimal choices of $\mathbf{Q}$ and $\Gen$ are
found using the method described in~\cite{Bergner:2011}.  A coarse
design of $19$ points is found in Figure \ref{f:designs}(a).  The
model runs represented by these points would be the first to perform
in the simulation and become available for intermediate data analysis.
The lattice is refined by adding another $18$ points (the number of
simulations is almost doubled, but one of the design points in the
refined lattice fell outside of $[0,1]^d$). This is done by applying
the inverse of a dilation matrix to the coarse lattice (alternatively,
the dilation matrix could be applied to the more dense lattice to
obtain the coarse lattice).  The combined design is shown in Figure
\ref{f:designs}(b). Notice that both the initial lattice and the
refined lattice are rotated and scaled versions of the same lattice.
Furthermore, the shape of the Voronoi regions are the same, and the
volume of the refined cells (area in 2-$d$) is one-half the volume of
the coarse cells.

The designs that we performed in the 8-dimensional input region above
were constructed in exactly the same manner.  That is, an initial
lattice design was found for a run size of $n_1=25$.  The next
$n_2=27$ points were found by applying the inverse of a dilation
matrix to the lattice and using the points that intersection the
8-dimensional unit hypercube.  Finally, the last $n_3=47$ points were
found in exactly the same way.

To summarize, space-filling design sequences can be obtained by
extending a good initial lattice design to a coarsening or refining
sequence of lattices. The rotational symmetry of the sequences
developed here ensures directionally uniform density at each level of
refinement.

\subsection{Cosmological Considerations}

Currently, observational constraints on neutrino masses are only upper
bounds, and in the Standard Model of Cosmology we still have
$m_\nu=0$. From a design stand-point, this puts the currently best-fit
value for one of our parameters at the edge of the design (since
centering the neutrino mass range around zero is excluded).  In
general, the accuracy of the emulator predictions start to degrade
when they approach a design edge, due to lack of nearby points. In
order to mitigate this problem, we add a set of 10 simulations with
$m_\nu=0$ to the design. We choose the values for the remaining 7
parameters according to a symmetric latin hypercube design.

While checking the accuracy properties of a first test emulator, we
noticed that models with $w_0+w_a$ approaching zero were not predicted
very well. This was due to the fact that the behavior of these models
on large scales is different compared to models closer to the Standard
Model.  With the other six parameters fixed, the correlation between
two models with $w_0+w_a$ close to zero is lower than between two
models with $w_0+w_a$ far from zero.  Observationally, $w_0+w_a$ close
to zero is already disfavored and predictions in this regime do not
necessarily need to be accurate at the percent level. Nevertheless we
aim to provide reliable predictions over the full parameter range we
cover.  Rather than alter our emulation scheme, we introduce a
transformation of these two parameters to scale the distances in these
parameters in a way that produces stationary behavior.  After some
exploratory data analysis, we replaced $(w_0,w_a)$ in the design space
with $(w_0,[-w_0-w_a]^{1/4})$.  This transformation results in the
desired behavior.  This special treatment of $(w_0,w_a)$ pairs can be
seen in Figure~\ref{w0wa} -- the design provides slightly more points
close to the limit $w_0+w_a=0$.

\section{$N$-body Simulations}
\label{nbody}

The parameter range outlined in Section~\ref{params} requires some
extensions of the standard setup for $w$CDM N-body simulations. In
particular, dynamical dark energy models defined by an equation of
state parametrized by $(w_0,w_a)$ have to be included as well as a
treatment of massive neutrinos.  In the following we will give a brief
description of our implementations of these additions in the $N$-body
code HACC (Hardware/Hybrid Accelerated Cosmology Code) described in
detail in \cite{habib14}.

In order to estimate the overall computational needs with regard to
box size, mass and force resolution, number of realizations, and other
considerations, we will also discuss results from a $\Lambda$CDM
simulation suite throughout the following sections. We give a brief
overview of the simulations here.

\subsection{Required Modifications in the $N$-body Code}

The modifications in the HACC code to include neutrinos and dynamical
dark energy models are straightforward and have been discussed in
\cite{upadhye14}.  In that work, HACC simulations were used to test
the range of validity of the TimeRG (``Renormalization Group'')
perturbation theory approach, introduced by \cite{timeRG1} and
\cite{timeRG2}, focusing on massive neutrinos and dynamical dark
energy models. This work will be an important ingredient for our
planned power spectrum emulator development.  We summarize the main
points here and remind the reader of the approximate treatment of
massive neutrinos that we employ. The impact of dynamical dark energy
and neutrinos are taken into account by (i) modifying the initializer
and (ii) including both effects in the background evolution.

\subsubsection{Neutrinos}

Simulating neutrinos fully self-consistently as a separate particle
species in a cosmological simulation is nontrivial mainly due to two
reasons: (i) the very large neutrino velocities at early times due to
the thermal velocity contribution which make very small time steps
necessary, (ii) the large mass difference between the tracer particles
representing dark matter and neutrinos respectively, which can lead to
scattering artifacts if not treated properly. Over the years, many
different solutions have been suggested to overcome these problems,
from starting the (neutrino) simulation very late (leading to a
different set of problems and inaccuracies) to simulating the
neutrinos on a coarser grid to treating the neutrinos only
linearly. An incomplete list of papers employing different solutions
include \cite{klypin}, \cite{gardini}, \cite{brandbyge1},
\cite{brandbyge2}, \cite{brandbyge3}, \cite{viel}, \cite{bird}, and
\cite{agarwal}.

In our work, we follow closely the approach used in \cite{agarwal}
with some modifications. We do not include the nonlinear evolution of
the neutrinos but instead evolve the cold dark matter-baryon component
($cb$) only, including the neutrinos in the background evolution. In
order to obtain the total matter power spectrum, we add at any given
output the linear component of the neutrino power spectrum to the
nonlinear $cb$ matter power spectrum in the following way:
\begin{equation}
P_{total}^{nl}(k,a)=\left[\sqrt{P_{cb}^{nl}(k,a)}+\sqrt{P_\nu^{lin}(k,a)}\right]^2. 
\end{equation}
This treatment is justified since neutrinos are massive enough to
behave matter-like close to $z=0$ but light enough to not cluster
strongly on small scales. In~\cite{upadhye14} we investigated the
validity of this approximation on large to quasi-nonlinear scales and
found sufficiently small errors for neutrino masses below the
currently excluded value.  

When setting the initial conditions, we include the neutrino
contributions in the transfer function and generate a linear power
spectrum at $z=0$ normalized to a given $\sigma_8$. Then, the $cb$
component of the matter power spectrum is moved back to the initial
redshift with a scale-independent growth function (the growth function
that includes neutrinos would be scale-dependent).  This growth
function includes all the species in the homogenous background. Note
that using a scale-dependent growth function would be inconsistent
since the evolution to $z=0$ is carried out for the $cb$
component. The setup described here ensures that at $z=0$ the power
spectrum matches the full linear power spectrum on large
scales. Finally, in the evolution the massive neutrinos are included
in the background evolution equation, but we do not source the
neutrino growth in the Poisson equation.  For a detailed discussion of
the implementation as well as various tests, the reader is referred to
\cite{upadhye14}.

\subsubsection{Dynamical Dark Energy Equation of State}

Next, we discuss our treatment of a dynamical dark energy equation of
state.  Without a compelling theory for a dark energy model beyond the
cosmological constant, the simplest way to express a dynamical origin
for dark energy is to parameterize the dark energy equation of state
$w(z)$ via a time varying functional form described by two parameters
($w_0, w_a$). A commonly used form is that suggested by
\cite{chevalier} and \cite{linder}:
\begin{equation}\label{param}
w(a)=w_0+w_a(1-a).
\end{equation}
In order to account for the time dependent dark energy equation of
state in the initial conditions of the simulations, we need to
generate a transfer function that includes the effects from
$w(z)$. For this, we modify the publicly available version of
{\tt{CAMB}}~\citep{camb} and include the dynamical dark energy in the
evolution of the background as well as modify the equations describing
the density and velocity perturbations in the dark energy\footnote{The
  modified {\tt CAMB} version can be downloaded at:\\
  http://www.hep.anl.gov/cosmology/pert.html}. 

To modify the background cosmology, we simply have to change the
equation describing the evolution of conformal time as a function of
the scale factor. Modifying the equations describing the perturbations
of dark energy requires an expression for the speed of sound. In
analogy with a simple scalar field model, we assume that this is the
speed of light. As a result, perturbations of dark energy develop only
on the Hubble scale. A second issue is that the equations describing
the evolution of the velocity perturbations include terms of the form
$c_s^2/(1+w(a))$, where $c_s$ is the speed of sound in the rest frame
of dark energy. For those values of $(w_0,w_a)$ for which the equation
of state passes through $-1$, this results in a singularity.  We treat
this by assuming the microscopic properties of dark energy do not
modify the power spectra, and so this term is small when $w(a)$
crosses $-1$. Next, we need to include $(w_0, w_a)$ in the growth
function in order to set up the initial conditions at the desired
redshift. This is achieved by evaluating the integral over $w(a)$ that
appears in $H^2(a)$ in the growth function:
$-3\int^a_1da'[w(a')/a']=a^{-3(w_0+w_a)}\exp[-3w_a(1-a)]$.  Finally,
the modifications in the Vlasov-Poisson equation for the time
evolution to include $(w_0, w_a)$ are also simple. In the case of
HACC, the expansion factor $a$ is used as the time variable, and
therefore the only change again is in $H(z)$ and the corresponding
scale factor expressions.

\subsection{$\Lambda$CDM Simulation}
\label{sim}

In order to test our overall simulation strategy with regard to
resolution and sufficient statistics, we carry out two high-resolution
$\Lambda$CDM simulations. These simulations represent the
specification of the main simulation suite that will be carried out
for each model in our future work. In addition to these high mass and
force resolution simulations, we will augment each model with a set of
lower resolution simulations to cover the full dynamic range with
sufficient accuracy as required. The cosmological parameters are
chosen to be close to the best-fit WMAP-7
parameters~(\citealt{wmap7}):
$\omega_m=0.1335,~\omega_b=0.02258,~n_s=0.963~,
w=-1.0,~\sigma_8=0.8~,h=0.71$. In one simulation we evolve 3200$^3$
particles in a (2100~Mpc)$^3$ volume, leading to a mass resolution of
$m_p\sim 10^{10}$M$_\odot$.  The force resolution is chosen to be
$\sim$6.6~kpc. This simulation has been previously used to validate
the accuracy of a $w$CDM power spectrum emulator~\citep{emu_ext} and
to create an emulator for the galaxy power spectrum~\citep{kwan13}
based on halo occupation distribution (HOD) modeling. The idea behind
using an HOD to generate galaxy catalogs is straightforward
(\citealt{kauffmann97}; \citealt{jing98}; \citealt{benson00};
\citealt{peacock00}; \citealt{seljak00}; \citealt{berlind02};
\citealt{zheng05}). The HOD provides a prescription for the galaxy
population (central and satellites) for a halo as a function of halo
mass.  The HOD parameters are obtained from matching the resulting
galaxy correlation function to observations. For different types of
galaxies, the HOD will be specified by different parameter values.
The resolution of our simulations is sufficient for a wide range of
investigations and to build detailed synthetic skies in different
wavebands.

The specifications of the additional simulations needed to generate
accurate emulators depend on the statistics of interest. As we show
below, for the power spectrum predictions we require a set of
particle-mesh (PM) runs for the quasi-linear regime to suppress the
noise due to realization scatter. For the mass function, we require
large volume simulations with high force resolution in addition to the
simulation described above to cover the cluster mass range.  As we
show in Section~\ref{massf}, a simulation covering a volume of
(5600~Mpc)$^3$ evolving 4096$^3$ particles is optimal for this
purpose. Due to the modest mass resolution of $\sim 10^{11}$M$_\odot$
and -- as we show below -- the rather small number of time steps
needed to achieve sufficient accuracy, the cost of these large
simulations is reasonable. A summary of the $\Lambda$CDM simulations
used in this paper is given in Table~\ref{tab_sum}.

\begin{table}[t]
\caption{Simulation specifications\label{tab_sum}}
\begin{tabular}{cccc}
$n_p$ & $L$[Mpc] & Realizations & Force resolution [kpc] \\
\hline\hline 
3200$^3$ & 2100 & 1 & 6.6 (high-res)\\
4096$^3$ & 5634 & 1 & 13.8 (high-res)\\
512$^3$ & 1300 & 16 & 1270 (PM) \\
\end{tabular}
\end{table}

\section{The Power Spectrum}
\label{power}

\subsection{Smooth Power Spectrum Predictions}

\begin{figure}[t]
\centerline{\hspace{-0.2cm}\includegraphics[width=3.35in]{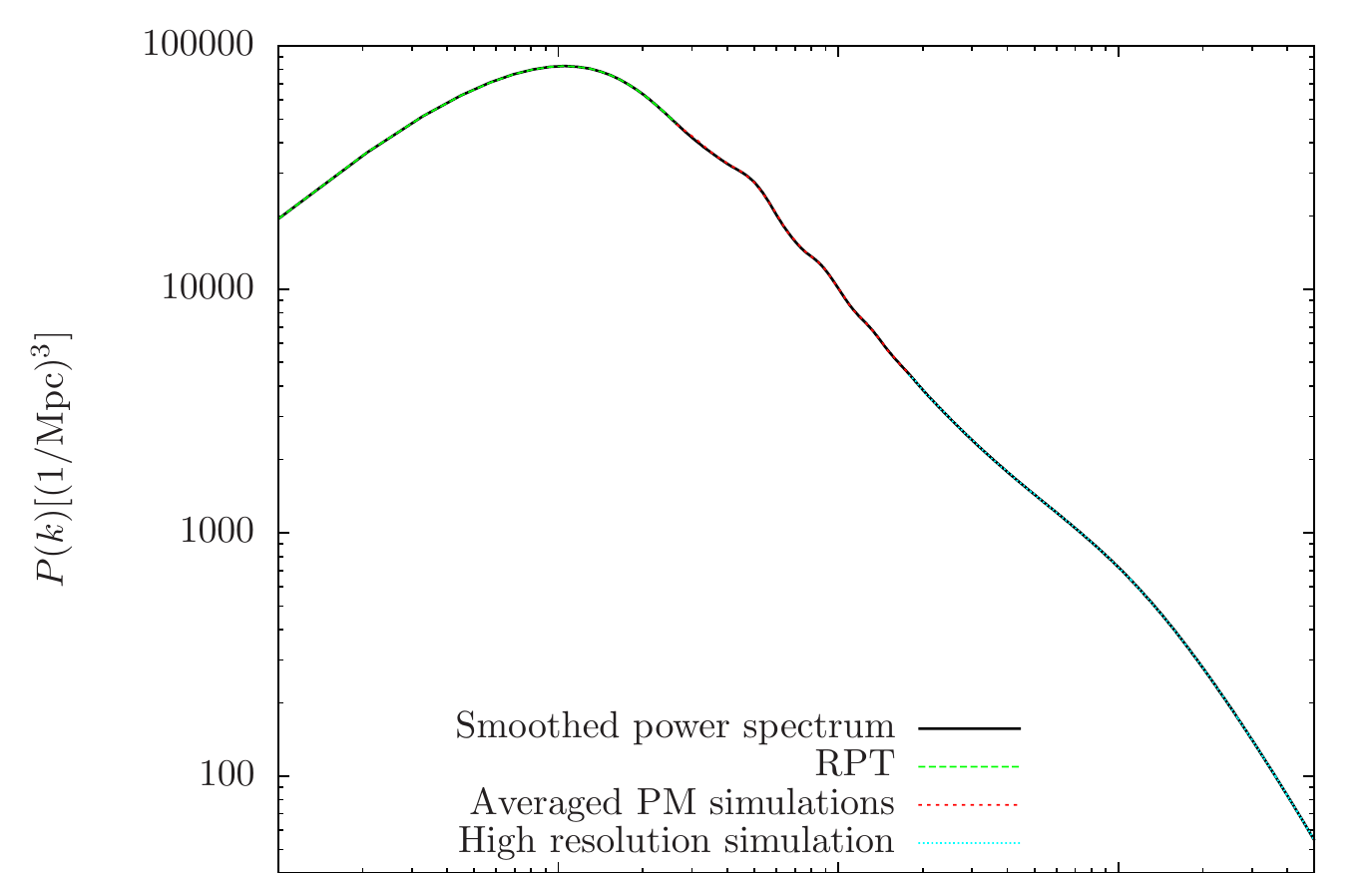}}
\centerline{\includegraphics[width=3.3in]{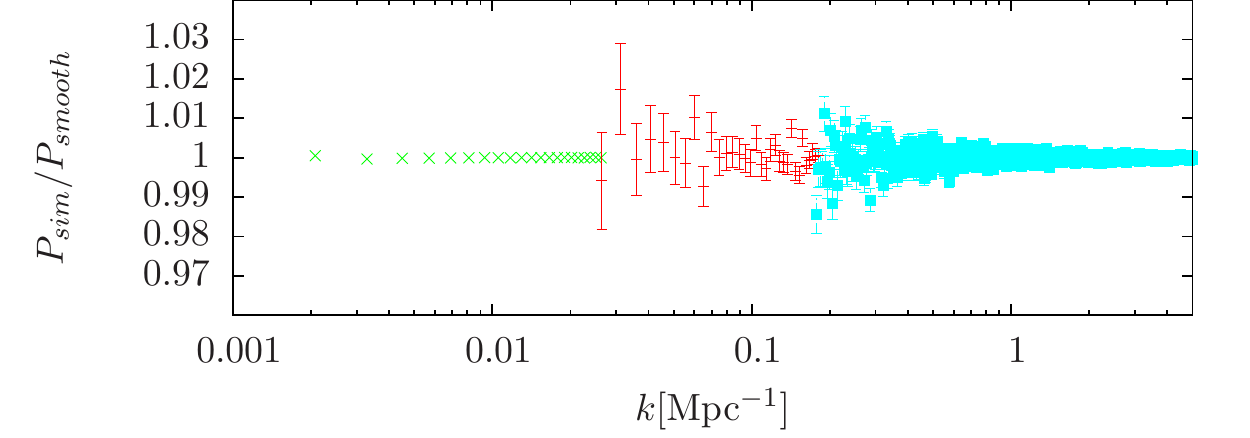}}
\caption{\label{pow_m000}Upper panel: Power spectrum at $z=0$
  assembled from RPT (green dashed), the average of 16 PM simulations
  (red dashed), and one high resolution simulation at large scales
  (light blue dotted). The black curve underneath the
  simulation/perturbation theory results shows the smooth power
  spectrum extracted from this ensemble. In order to demonstrate the
  quality of the smoothing procedure, we show the ratio of these
  results in the lower panel, including error bars calculated from
  the finite number of modes at each bin for the simulation results.}
\end{figure}

Following the discussions in \cite{coyote3} and \cite{emu_ext}, a
major obstacle in extracting a smooth power spectrum from a simulation
campaign is due to realization scatter.  In order to solve this
problem, we have shown that a smoothed power spectrum can be achieved
by matching higher order perturbation theory on the largest scales
with medium resolution, particle mesh (PM) runs (many realizations) on
intermediate scales, and a high resolution simulation on small
scales. In addition to this matching procedure, we have developed a
sophisticated smoothing method based on a process convolution approach
described in~\cite{higdon2002}. The main idea behind this approach is
to build a smooth function as a moving average of a simple stochastic
process. The moving average uses a smoothing kernel with a varying
width -- in regions with distinct features like the baryon acoustic
oscillations the window is narrow while in very smooth regions such
the very small scales, the window can be wider. We will use the same
method again here. We cover a $k$ range between 0.001 to 6~Mpc$^{-1}$
via matching (i) renormalized perturbation theory out to
$k\sim0.04$~Mpc$^{-1}$ for $z=0,~1$ and $k\sim0.14$~Mpc$^{-1}$ for
$z=2$; (ii) 16 realizations of PM simulations with 512$^3$ particles
on a 1024$^3$ grid in a (1300~Mpc)$^3$ volume out to
$k\sim0.25$~Mpc$^{-1}$; (iii) a (2100Mpc)$^3$ volume, high-resolution
simulation with 3200$^3$ particles and 6.6~kpc force resolution to
cover the remaining $k$-range.

Figure~\ref{pow_m000} demonstrates the matching and smoothing strategy
for our fiducial $\Lambda$CDM model, M000, described in
Section~\ref{sim}. The power spectrum was obtained from the simulation
in the same way as described in \cite{coyote1} via mapping the
particles onto a large grid (6400$^3$ in this case) using a
cloud-in-cell (CIC) assignment, applying an FFT, correcting for the
charge-assignment window function and averaging the results in bins in
$|{\bf k}|$. We show results for both $P(k)$ and the dimensionless
power spectrum $\Delta^2(k)$, connected via the following relation:
\begin{equation}
\Delta^2(k)=\frac{k^3P(k)}{2\pi^2}.
\end{equation}
We only present results at $z=0$ here, for $z=1$ and $z=2$ the reader
is referred to the Appendix of~\cite{emu_ext}. The upper panel shows
the full power spectrum assembled from RPT, PM simulations, and one
high-resolution run on top of the smoothed power spectrum. On the
logarithmic scales used here, even without the smoothing, the result
is already very good. The lower panel shows the ratio of the smoothed
power spectrum to the simulations; the aim of this figure is to
demonstrate that the matching does not lead to any bias.
These results show that our matching and smoothing strategy works as
desired. We also point out that the large simulation volume paired
with good mass resolution allows us to push out to higher $k$ without
having to rely on small simulation volumes, which can induce some
inaccuracy, as was shown in ~\cite{emu_ext}.

\subsection{Emulator Performance}

\begin{figure}[t]
\centerline{
 \includegraphics[width=3.in]{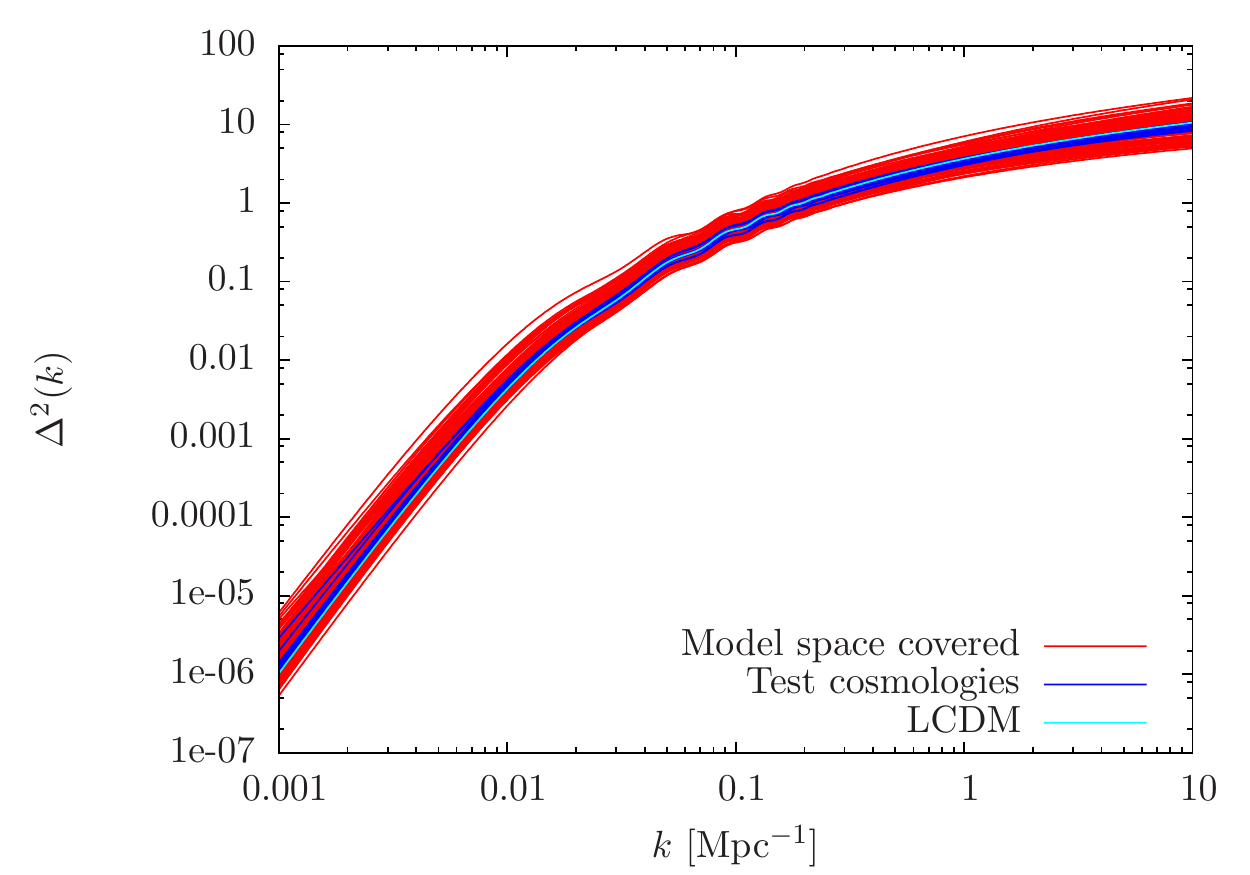}}
\caption{\label{delta}Power spectrum model space covered by 101
  cosmological models within the range described in
  Eqs.~(\ref{cosmoparams}), shown in red.  Test models are shown in
  blue, where the light blue curve shows M000, a $\Lambda$CDM model,
  and the dark blue curves show the additional models specified in
  Table~\ref{tab2}.}
\end{figure}

\begin{figure}[t]
\center{\includegraphics[width=1.7in]{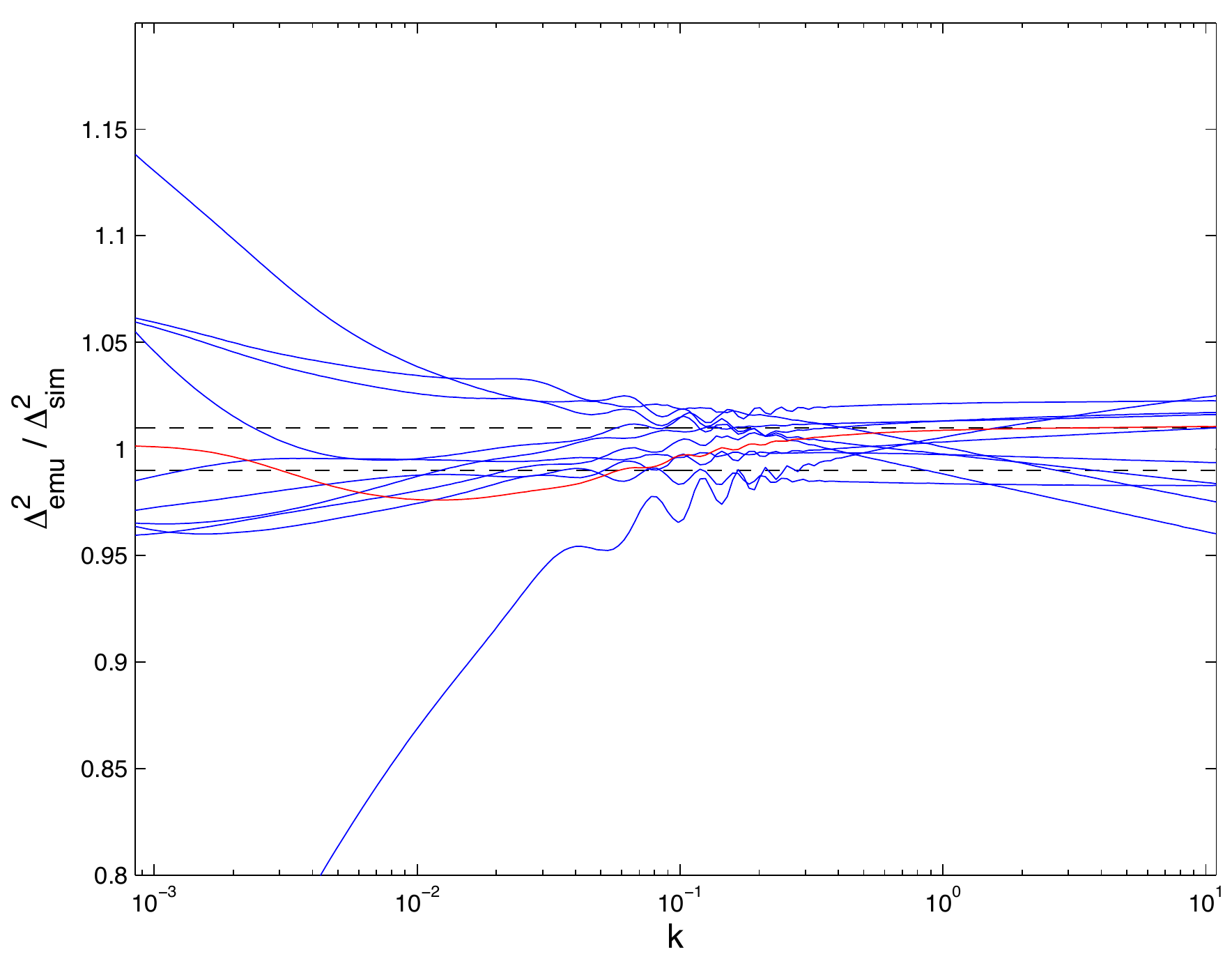} 
\includegraphics[width=1.7in]{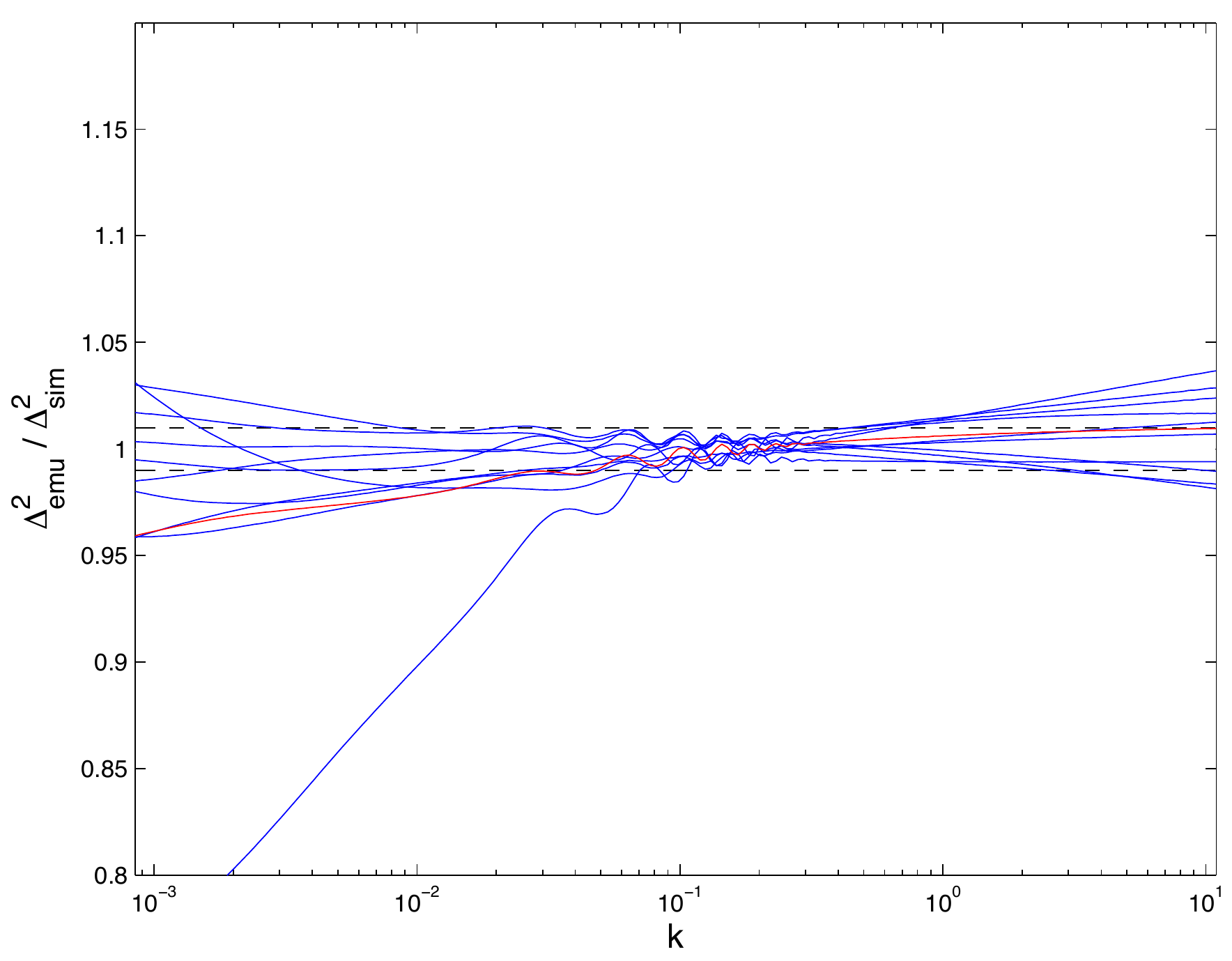}  
\includegraphics[width=1.7in]{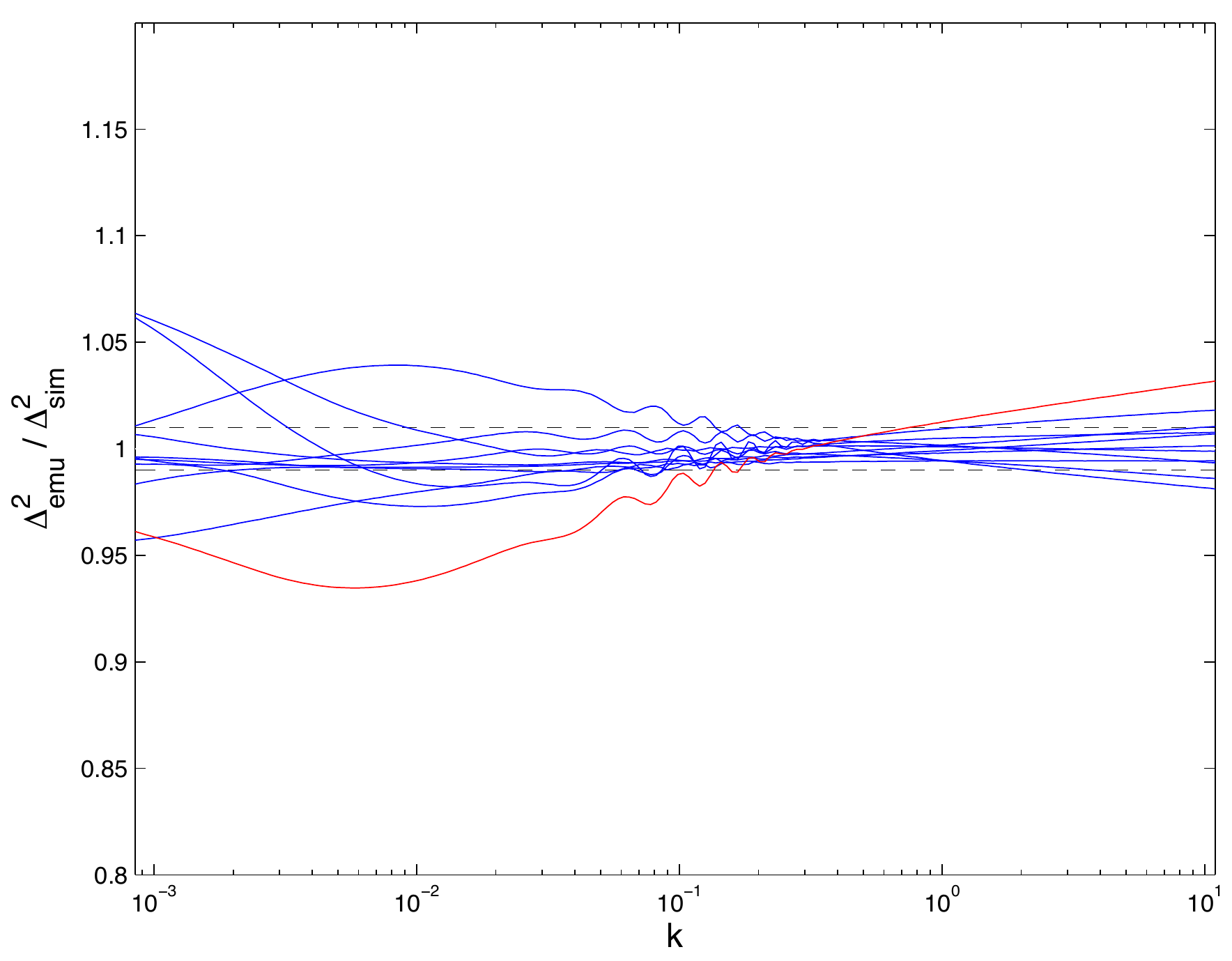}
\includegraphics[width=1.7in]{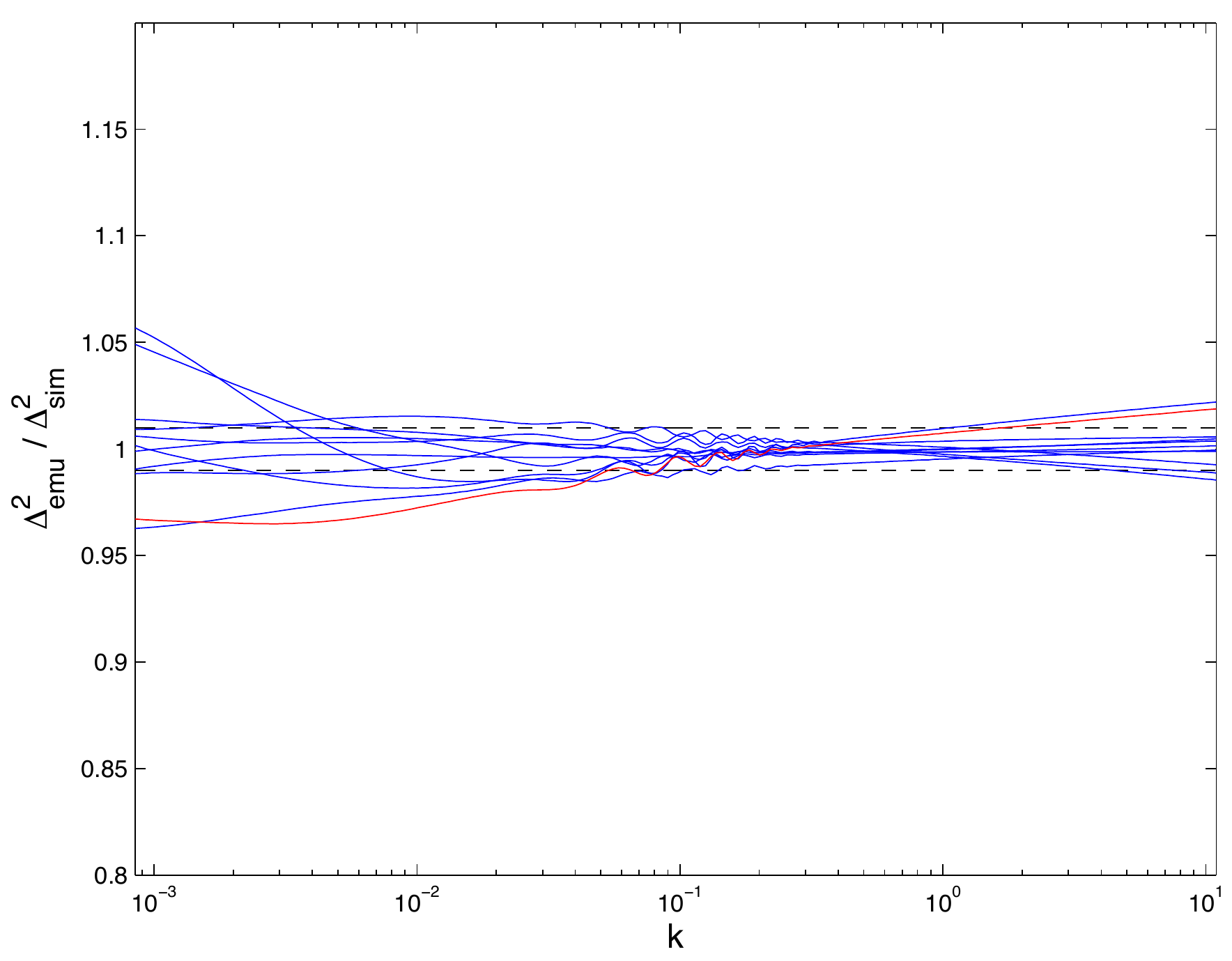}
\includegraphics[width=1.7in]{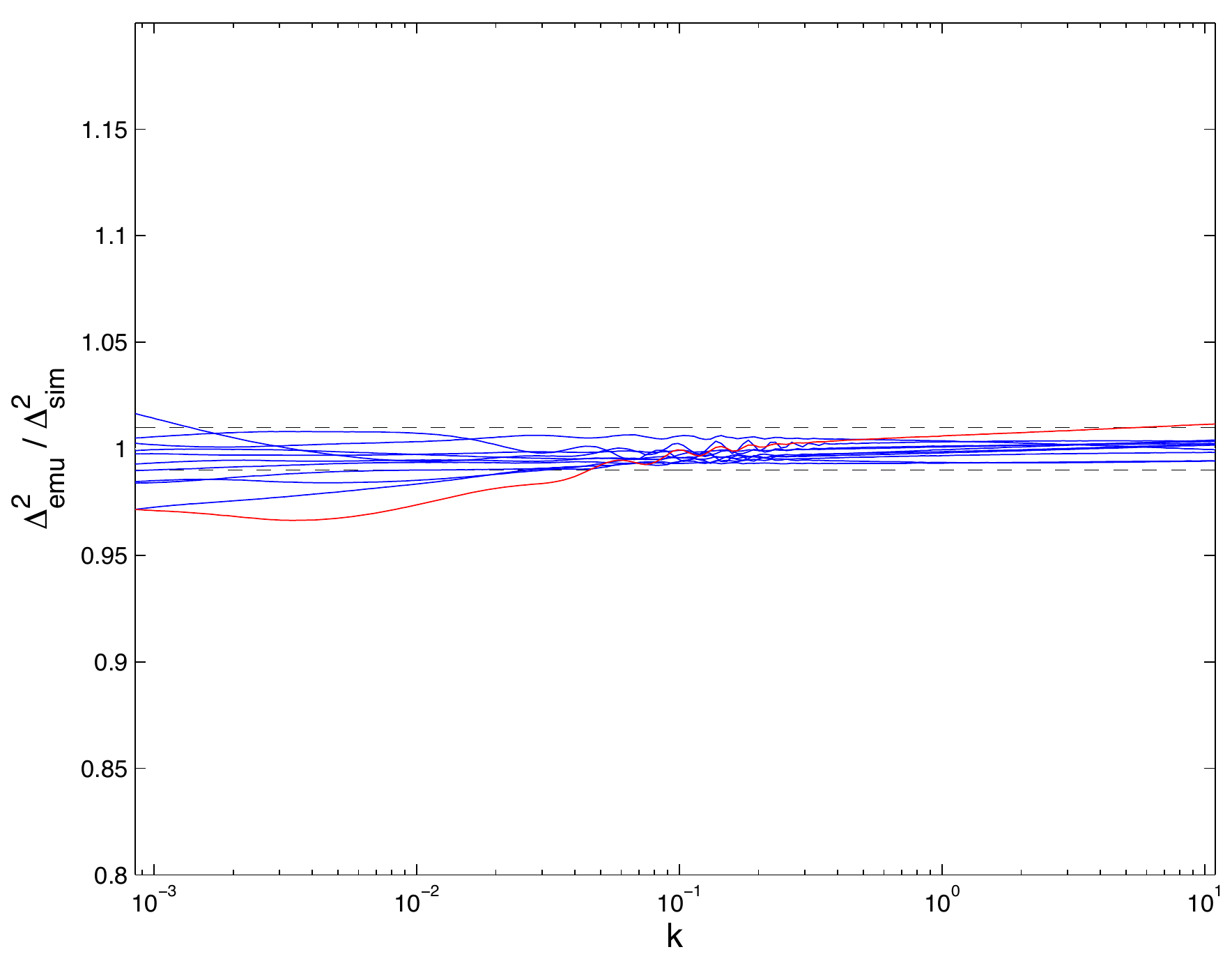}
\includegraphics[width=1.7in]{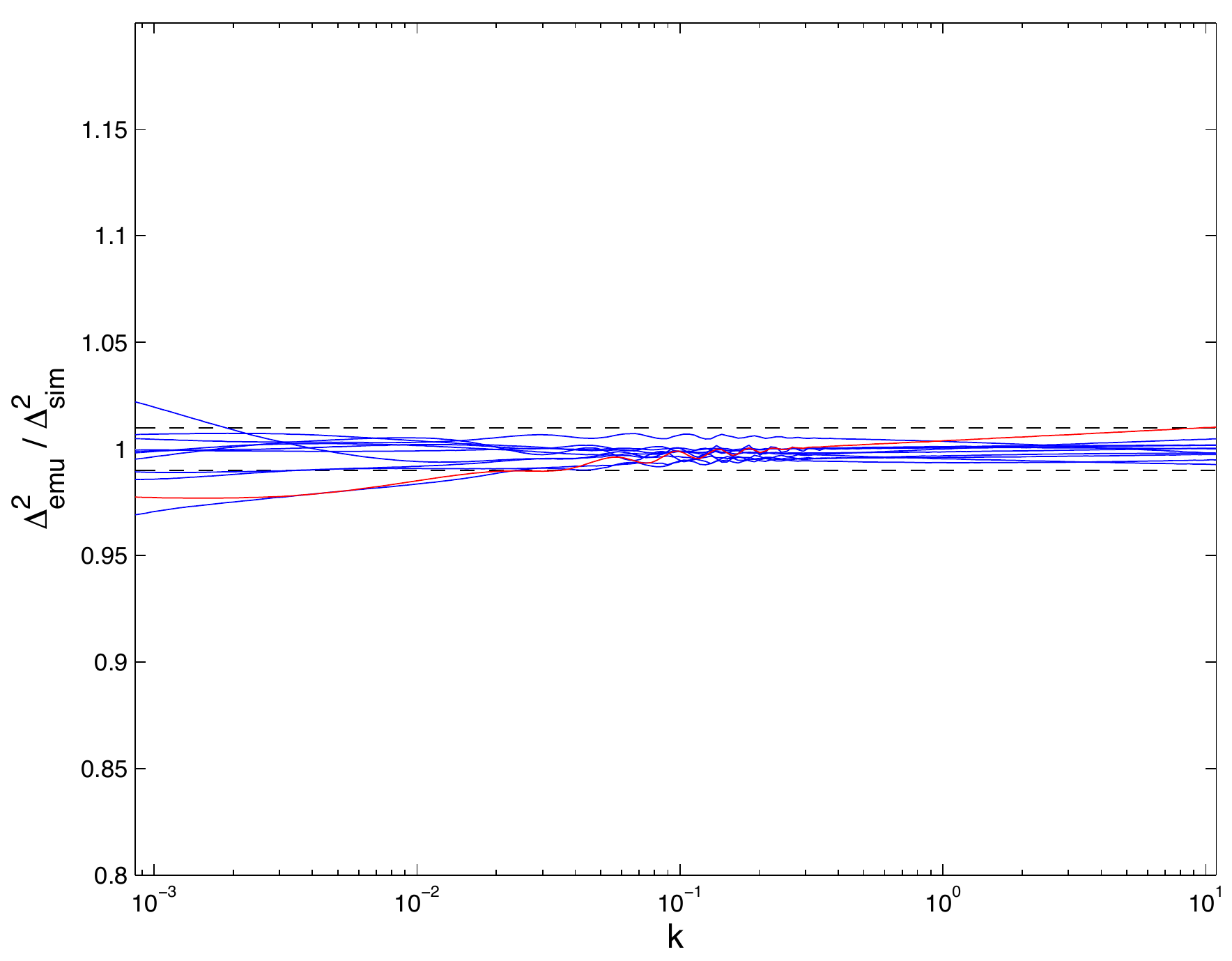}}
\caption{\label{emu_test}Emulator test: An emulator for the linear
  power spectrum is tested using 10 reference cosmological models. The
  emulator predictions built from up to 101 models are compared to the
  exact linear spectra from CAMB in the left panels and predictions
  from up to 111 models in the right panels. Shown is the ratio of
  prediction to ``truth'', the emulator predictions are accurate at
  the 1\% level (dashed line) on small scales. The upper panels shows
  the prediction for an emulator based on 26/36 models, the middle
  panel is based on 55/65 models, and for the predictions shown in the
  lower panel all 101/111 models were used. The red line shows the
  prediction accuracy for a $\Lambda$CDM model.}
\end{figure}

Following the strategy outlined in Section~\ref{design} we now discuss
the accuracy we expect to achieve from 101 simulation models, obtained
in a three-step simulation campaign. The first set of power spectra
covers 26 models, the second set 55, and the final set covers all 101
models. In addition, as discussed in Section~\ref{design} we add 10
models with $m_\nu=0$ to properly cover this important edge of the
design hypercube.

For our tests, we use linear power spectrum predictions generated with
{\tt CAMB}. This allows us to obtain many power spectra with nominal
computational effort. At the same time, the linear power spectrum has
been proven to be a very good testbed for the final nonlinear emulator
(see, e.g.,~\citealt{coyote2}). Figure~\ref{delta} shows the model
space we cover via the cosmological parameter ranges given in
Eqs.~(\ref{cosmoparams}) and the cosmological models used to test the
accuracy of the emulator. The model specifications are given in
Table~\ref{tab2} in Appendix~\ref{appendixa} as models E001 to E010.
These extra models were chosen by generating another design within the
parameter hypercube.  We do not use a model on the edge of the
hypercube for our test. At the edges, the accuracy requirements are
not as stringent as those models are observationally already ruled
out. Nevertheless, for the purpose of an inverse analysis using Markov
chain Monte Carlo (MCMC) methods, predictions over a wide parameter
space are desirable.

Results from the accuracy test are displayed in
Figure~\ref{emu_test}. We show the error of the emulators obtained
from 26, 55, and 101 models in the left three panels and the results
including the extra 10 runs for $m_\nu=0$ in the three panels on the
right. The accuracy improves considerably on adding more cosmological
models particularly on small scales. Considering the large parameter
space we cover, the accuracy obtained from only 26 models is already
extremely good. The two models that are not predicted well on large
scales (out of 26) are E007 and E009. For E007, we have
$w_0+w_a=-0.011$ and for E009 we have $w_0+w_a=-0.0222$.  Both of
these models therefore fall in the class of so-called ``early dark
energy models''.  The reason for the inaccurate prediction is a slight
increase in power on large scales in these models.  Nevertheless,
after increasing the model space to 56 and 101 the results of these
models are also accurately captured. Finally, in addition to the ten
models we also show the prediction for the $\Lambda$CDM case in
red. Here one can see how the additional 10 models with zero neutrino
mass help to improve the prediction accuracy. Overall, with the
complete set of models, we are able to provide predictions at the $\pm
1\%$ level (shown by the dashed lines in all panels) over all 8
parameters out to $k\sim 10$Mpc$^{-1}$. Extending the range to higher
$k$ values involves running higher-resolution simulations (both force
and mass) and properly accounting for baryonic effects. As discussed
in the Introduction, results from initial investigations are becoming
available for the latter. In the future, a combination of
extrapolation to large $k$ values and modeling of baryonic effects
will be required for robust predictions on the smaller length scales
relevant to cosmological analyses.

\section{The Mass Function}
\label{massf}

\subsection{Smooth Mass Function Prediction}

In the same way as for the power spectrum, building a mass function
emulator requires a smooth prediction for the mass function for each
cosmological model in our simulation design. For the mass function,
this task is much easier than for the power spectrum, since no subtle
features such as baryonic wiggles are present. The much more
challenging aspect here is to obtain enough statistics with regard to
halo counts; as has been shown previously (see, e.g.,
\citealt{lukic07,tinker08,crocce09,bhattacharya11}) capturing the mass
function accurately at the cluster mass scale is not easy. This is due
to the exponential drop-off of the mass function at large mass. Small
uncertainties in the mass estimates for massive halos are amplified and
lead to effects at the level of tens of percent in the mass function
itself. In order to generate an accurate mass function from
simulations, it is important to have (i) a large enough volume to
capture a large number of cluster sized halos, (ii) good force
resolution to ensure that the halo masses are accurate, and (iii) good
mass resolution to ensure that halos are sufficiently resolved to
ensure accurate mass determinations.

\begin{figure}[t]
\centerline{\hspace{-0.37cm}\includegraphics[width=3.35in]{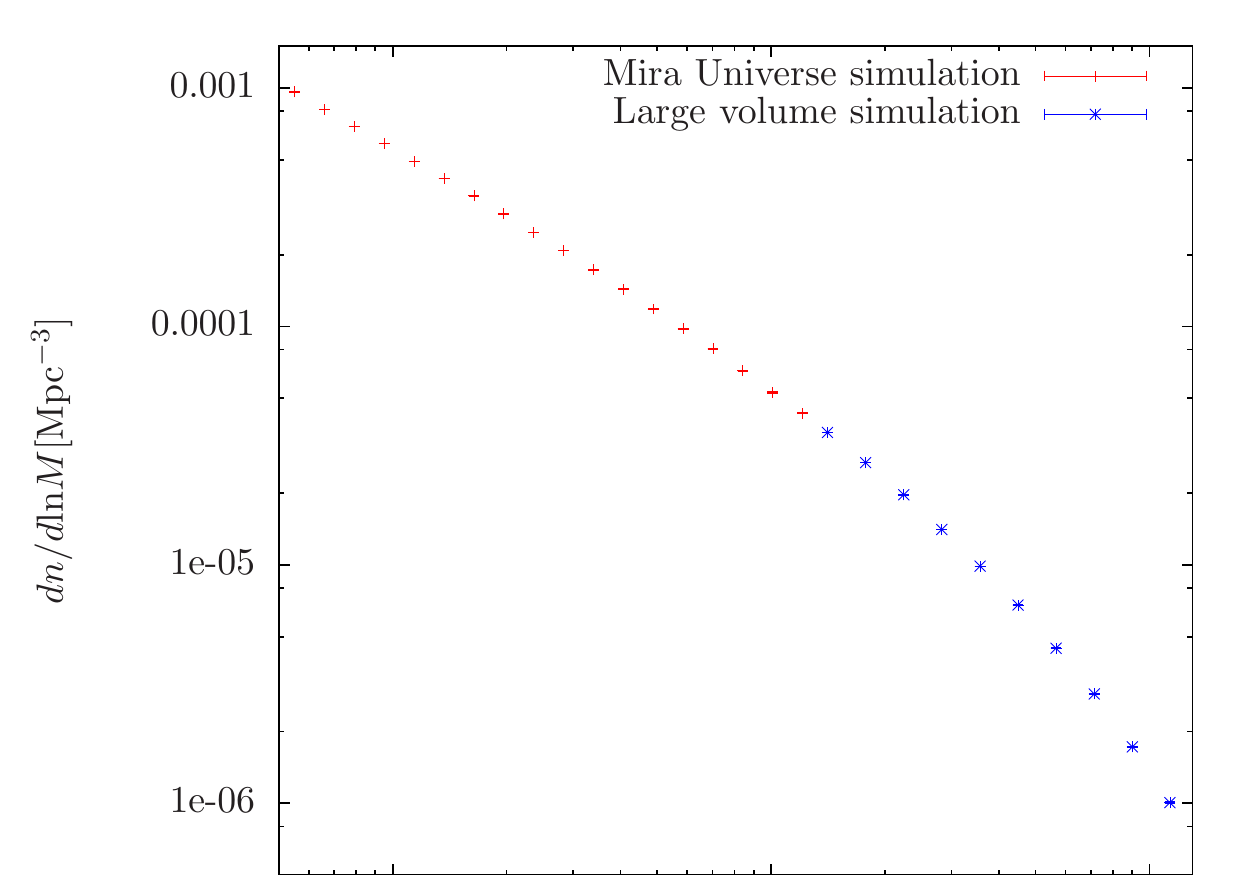}}
\centerline{\includegraphics[width=3.19in]{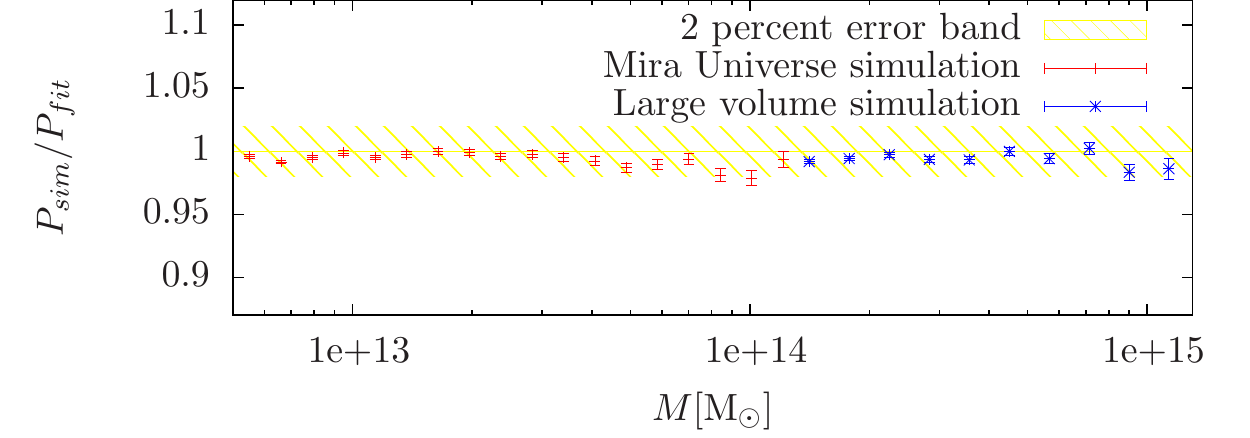}}
\caption{\label{massf_m000}Upper panel: Mass function including
  Poisson error bars from the Mira-Titan Universe simulation (red) to
  cover the low mass end of the mass function and an additional large
  volume simulation, spanning a (5664 Mpc)$^3$ volume (blue) to cover
  the cluster regime. Lower panel: Ratio of the simulation results
  with respect to the \cite{bhattacharya11} fit, assuming
  universality. The yellow shaded region shows the 2\% error band. }
\end{figure}

The mass range of interest for cosmological surveys in the area of
``cluster cosmology'' actually covers the range from galaxy groups to
large clusters. With a mass resolution of $m_p\sim1.05\cdot
10^{10}$M$_\odot$ for the $\Lambda$CDM case, the Mira-Titan Universe
comfortably resolves this range of masses. With a volume of
(2100Mpc)$^3$ the number of very massive clusters is not very large,
and we therefore limit the use of the Mira-Titan Universe run to a
mass range between $m_{\rm halo}\sim 5\cdot 10^{12}$M$_\odot$ to
$m_{\rm halo}\sim 1\cdot 10^{14}$M$_\odot$.  For the small mass end,
halos are therefore resolved with $\sim 500$ particles which minimizes
the need for (FOF) mass corrections due to undersampling of the
halos. For the high-mass halos, this cut ensures that we have tens of
thousands of halos in the relevant mass bins. In order to cover the
upper end of the mass function, we employ another set of
simulations. These simulations cover volumes of $\sim$ 5600Mpc$^3$ and
evolve 4096$^3$ particles, leading to a mass resolution of
$m_p\sim10^{11}$M$_\odot$. These simulations were designed to generate
mock catalogs for surveys such as BOSS and DESI~(Sunayama et al. in
preparation). At masses of $m_{\rm halo}\sim 1\cdot 10^{14}$M$_\odot$
halos are resolved with $\sim 1000$ particles.  If we restrict the
mass range to $m_{\rm halo}\sim 10^{15}$M$_\odot$, the last mass bin
has more 10,000 halos, leading to small statistical errors.

Figure~\ref{massf_m000} shows the results for the mass function from
the Mira-Titan Universe simulation for the lower mass end and the
results from the large simulation for the cluster regime.  We use the
friends-of-friends (FOF) definition for determining the halo masses
using a linking length of $b=0.2$~(\citealt{davis85}). This mass
function definition has been studied very thoroughly in the past and
allows us to compare our results to other work.  (Our discussion can
easily be extended to overdensity masses as well.)  The lower panel
displays the ratio of the simulation results with respect to a mass
function fit developed in \cite{bhattacharya11}. The simulations agree
with this fit to better than 2\% (yellow band) over the full mass
range.

\begin{figure}[t]
\centerline{\hspace{-0.cm}\includegraphics[width=3.35in]{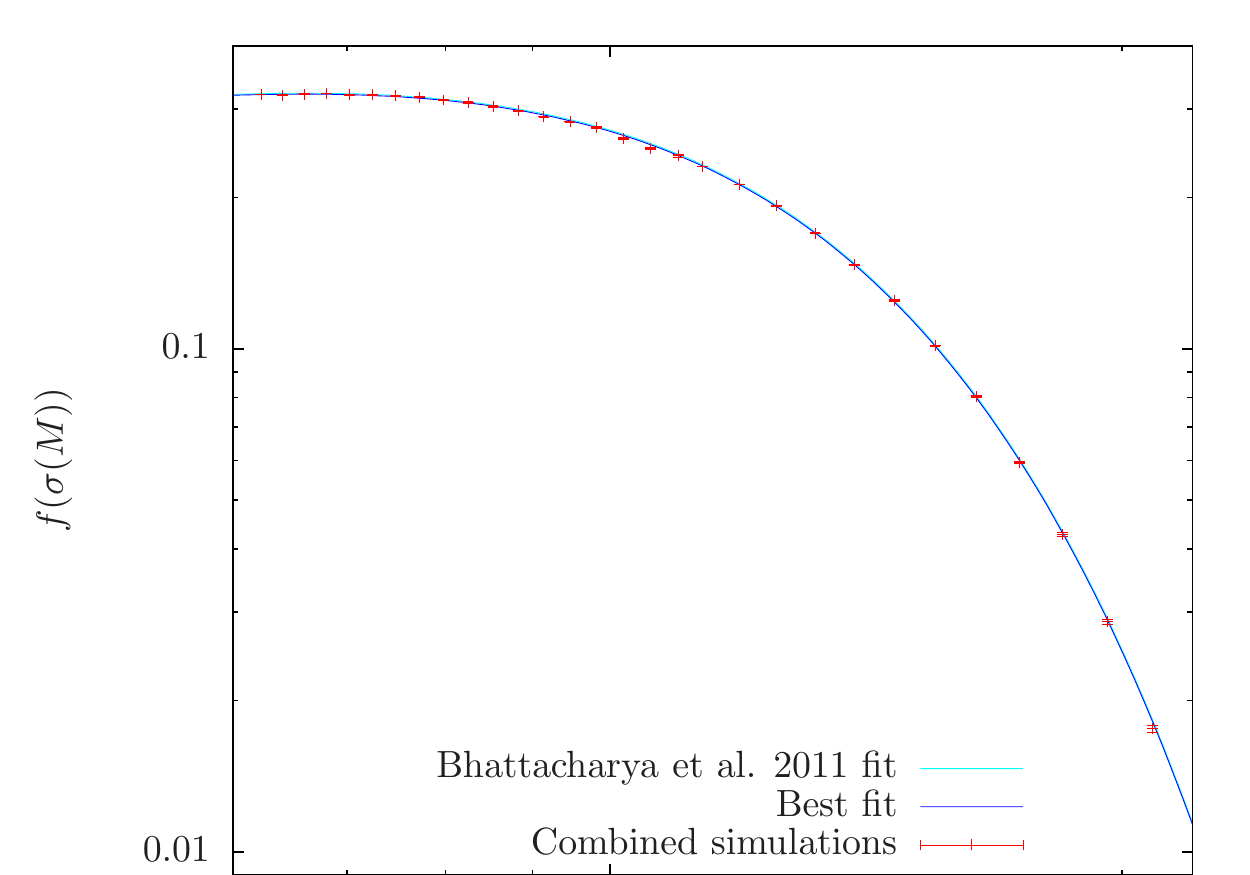}}
\centerline{\includegraphics[width=3.35in]{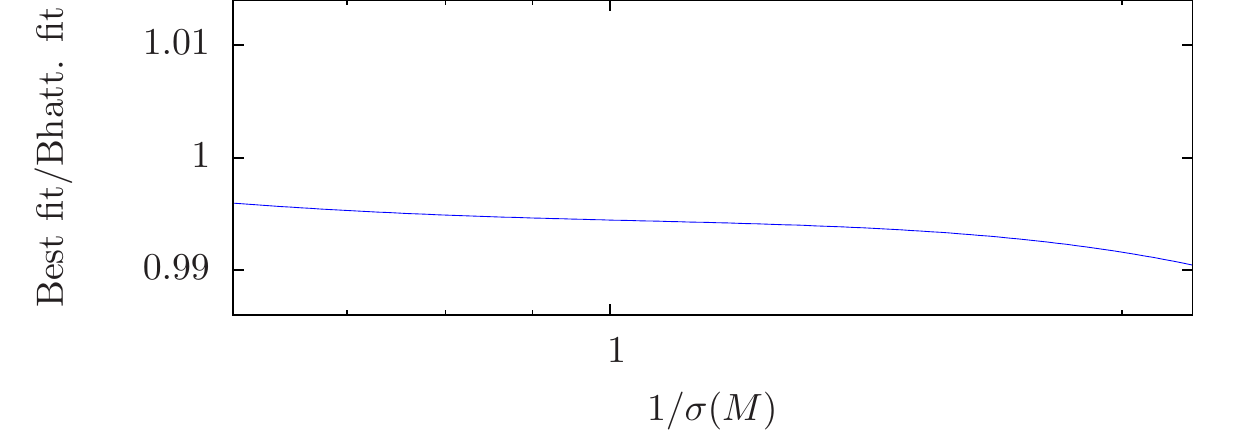}}
\caption{\label{sigma_m000}Upper panel: Combined simulation result
  (red) for $f(\sigma)$ compared to the best fit from
  \cite{bhattacharya11} and the best fit to the simulation result.  
  The two fits are basically indistinguishable. Lower panel: Ratio
  between the two fits. The agreement is better than one percent.  }
\end{figure}

The functional form of the \cite{bhattacharya11} fit is inspired by
the original Sheth-Tormen mass function fit \citep{sheth99} but has
one additional parameter and allows for explicit redshift dependence: 
\begin{equation}\label{fit1}
f^{\rm Bhatt}(\sigma,z)=A\sqrt{\frac{2}{\pi}}
\exp\left[-\frac{a\delta_c^2}{2\sigma^2}\right]
\left[1+\left(\frac{\sigma^2}{a\delta_c^2}\right)^p\right]   
\left(\frac{\delta_c\sqrt{a}}{\sigma}\right)^q,
\end{equation} 
with the following parameters (note that all parameters are different
from the original Sheth-Tormen choices): 
\begin{equation}\label{fit2}
A=\frac{0.333}{(1+z)^{0.11}};~~a=\frac{0.788}{(1+z)^{0.01}};
~~p=0.807;~~q=1.795,  
\end{equation}
with $\delta_c=1.686$. Figure~\ref{sigma_m000} shows the results for
$f(\sigma(M))$ as a function of $1/\sigma(M)$ as measured by the
combined simulations and the fit from \cite{bhattacharya11}. 
We have refitted all parameters for $z=0$
using our current simulations and find extremely good agreement with
the previous results: 
\begin{eqnarray}\label{fitnew}
&&A=0.3315\pm0.0003;~~a=0.7895\pm0.0085;\\
&&p=0.8148\pm0.0451;~~q=1.8001\pm0.0425,\nonumber
\end{eqnarray}
shown in Figure~\ref{sigma_m000}. The agreement between the fits at
the sub-percent (the lower panel in Figure~\ref{sigma_m000} shows the
ratio of the two fits) is rather remarkable and reassuring, given the
fact that the simulations were carried out with different codes. Since
the simulation results are very smooth already, we do not need to
employ a special smoothing procedure. Instead, we will be able to
build the emulator directly from the binned mass function datasets,
eliminating one source of uncertainty in the emulator construction.

\begin{figure}[t]
\centerline{
 \includegraphics[width=3.in]{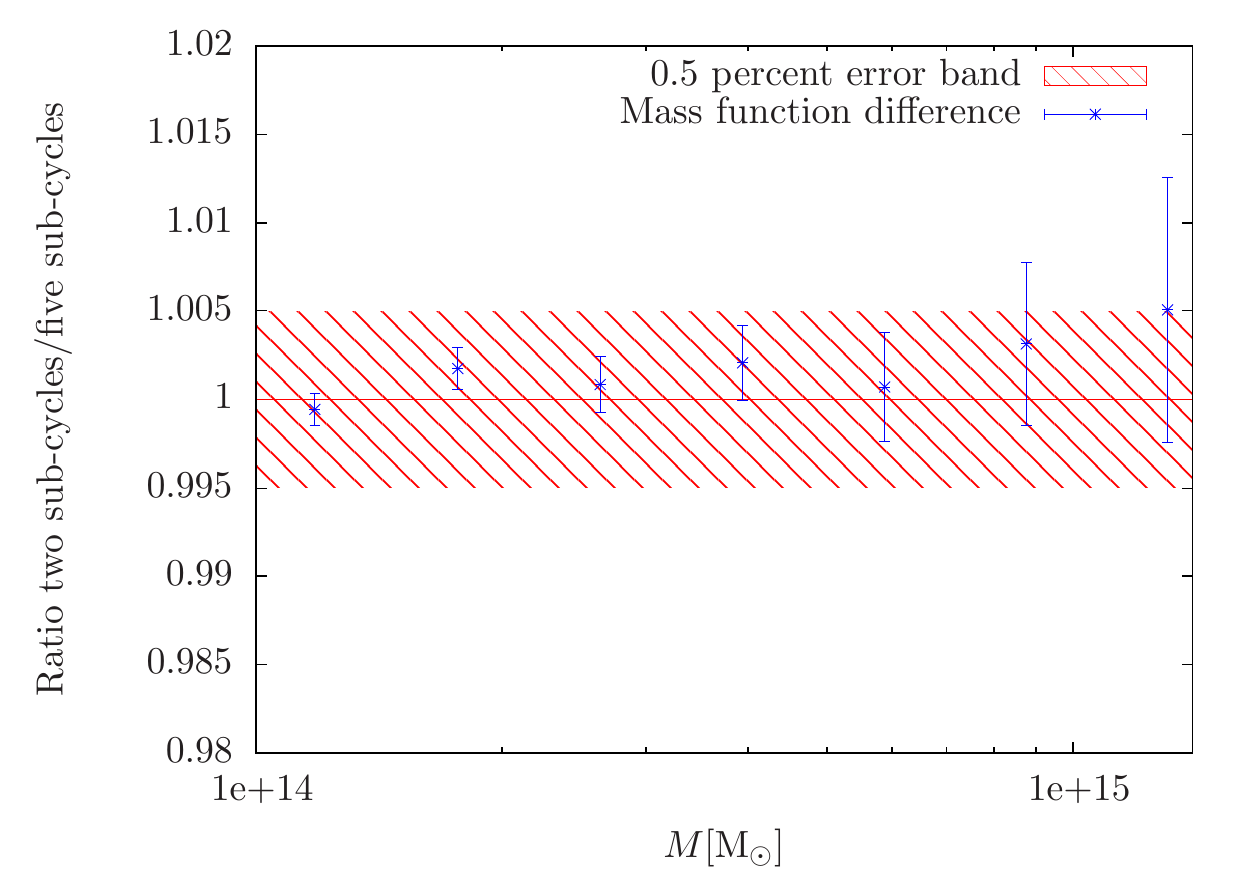}}
\caption{\label{time_step}Mass function error in the regime of cluster
  masses due to coarsening the time stepping. Shown is the ratio of a
  simulation with 2 sub-cycles per long-range step over 5
  sub-cycles. In the mass regime where the large volume runs will be
  used, the difference is at the sub-percent level. Given the
  computational saving of a factor of 2.5, this small
  inaccuracy is tolerable.}
\end{figure}

Next, we briefly discuss how to reduce the computational costs for the
large simulation volume runs. In Sunayama et al. (in preparation) a
detailed study is carried out concerning the number of time steps
needed to obtain accurate predictions for halo statistics.  In the
HACC code the time stepper is divided into long time steps for the
long-range force solver and each long time step is subdivided into
shorter time steps with short-range force solves in order to resolve
small scale structures accurately. At a mass resolution of $m_p\sim
10^{11}$M$_\odot$ and with the mass function as the observational
target, the number of sub-cycles can be reduced without losing
accuracy. Figure~\ref{time_step} shows the ratio of the mass function
in the mass regime of interest for the large simulation with 2 and 5
sub-cycles (earlier tests shown in, e.g.,~\cite{habib14} have
demonstrated convergence for 5 sub-cycles). The difference is at the
sub-percent level, with computational costs reduced by a factor of
2.5. At the same time, these simulations will be still very useful for
generating synthetic sky catalogs for large scale structure surveys
such as DESI.

Our final mass function emulator will go beyond a single mass
definition. We will measure mass functions from our simulations for
both FOF and overdensity definitions, varying linking length and
overdensity specification.  This will make it more convenient to
compare the results to observational mass measurements and proxies as
well as to study connections between the different halo
definitions. The extension of our results shown above to build a
flexible emulator like this is straightforward.

Finally, we comment on the higher cluster mass range of the mass
function.  In this paper, we have been very conservative with regard
to pushing the mass range too far. In future work, following the
discussion in \cite{lukic07}, we will include careful investigations
on finite volume effects in our simulations allowing us to push to
higher masses. We will provide a more detailed discussion on this
topic in a forthcoming paper, where we focus solely on the mass
function.

\subsection{Emulator Performance}

In order to test the emulator performance following the design
strategy outlined in Section~\ref{design}, we need a mass function
surrogate -- just as we used linear theory for the power
spectrum. First suggested in \cite{J01}, writing the mass function in
terms of $\sigma(M)$ allows for an almost universal description of the
friends-of-friends (FOF) mass function for a linking length of $b=0.2$
that depends only on the linear power spectrum. Over time this
universality has been investigated in detail (see, e.g.,
\citealt{reed06,cohn07,lukic07,
  tinker08,crocce09,courtin10,bhattacharya11}) with the conclusion
that for the specific definition of the $b=0.2$-FOF mass function,
universality holds at the 10\% level accuracy over a wide range of
cosmologies. We use this result in the following to investigate the
accuracy of a mass function emulator over the parameter space
discussed in this paper.

\begin{figure}[t]
\centerline{
 \includegraphics[width=3.in]{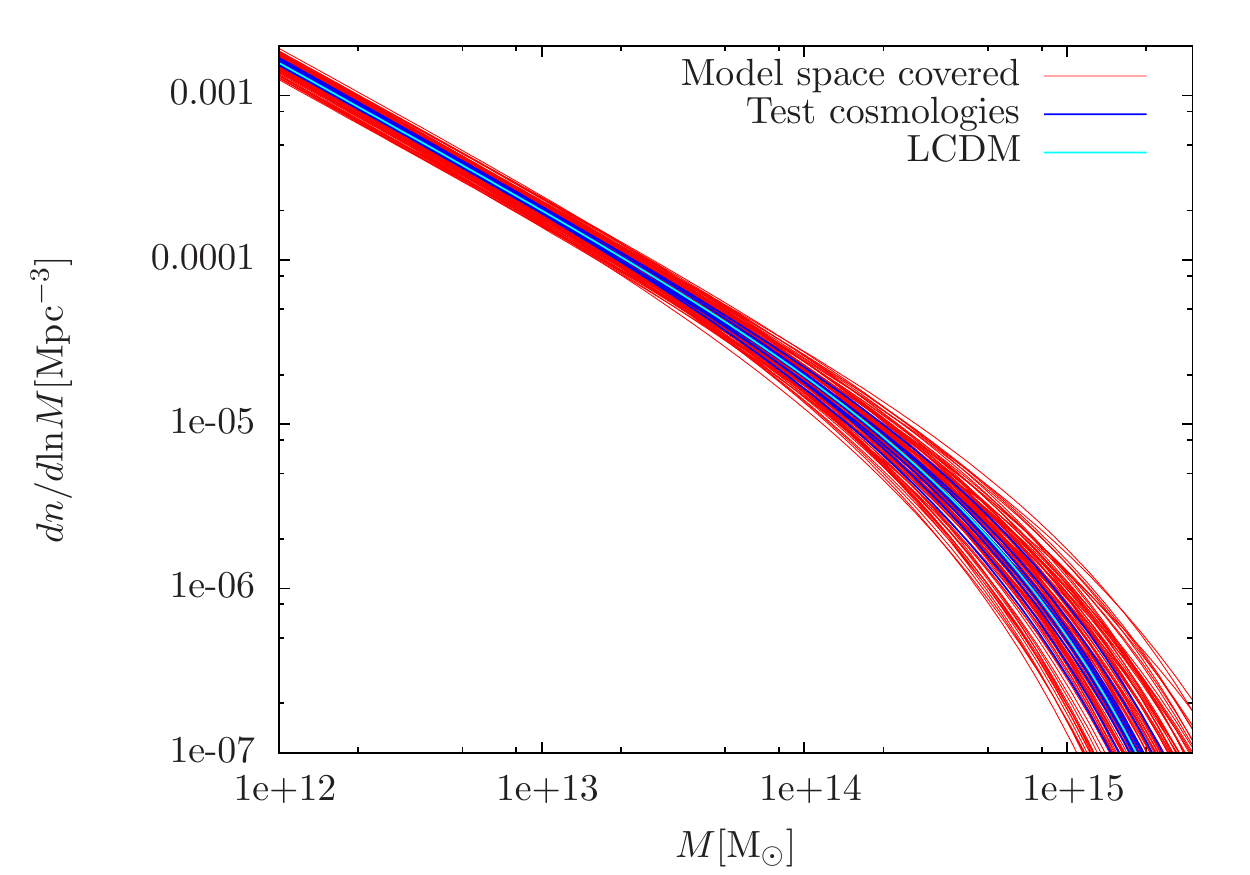}}
\caption{\label{massf_0}Coverage of mass function model space with the
  simulation design described in this paper at $z=0$ (red).  The dark
  blue lines show the test models and the light blue line the fiducial
  $\Lambda$CDM mass function. All models are based on assuming
  universality is valid over the full parameter range. This assumption
  is valid at the $\sim 10\%$ level of accuracy.}
\end{figure}

We use the universal form of the mass function given in
Eq.~(\ref{fit2}) by \cite{bhattacharya11} to generate mass function
predictions for all our models. Figure~\ref{massf_0} shows the model
space covered by the simulation campaign described in this paper as
well as the test cosmological models used to gauge the emulator
accuracy. The $\Lambda$CDM model used throughout the paper is also
shown.

\begin{figure}[t]
\center{\includegraphics[width=1.7in]{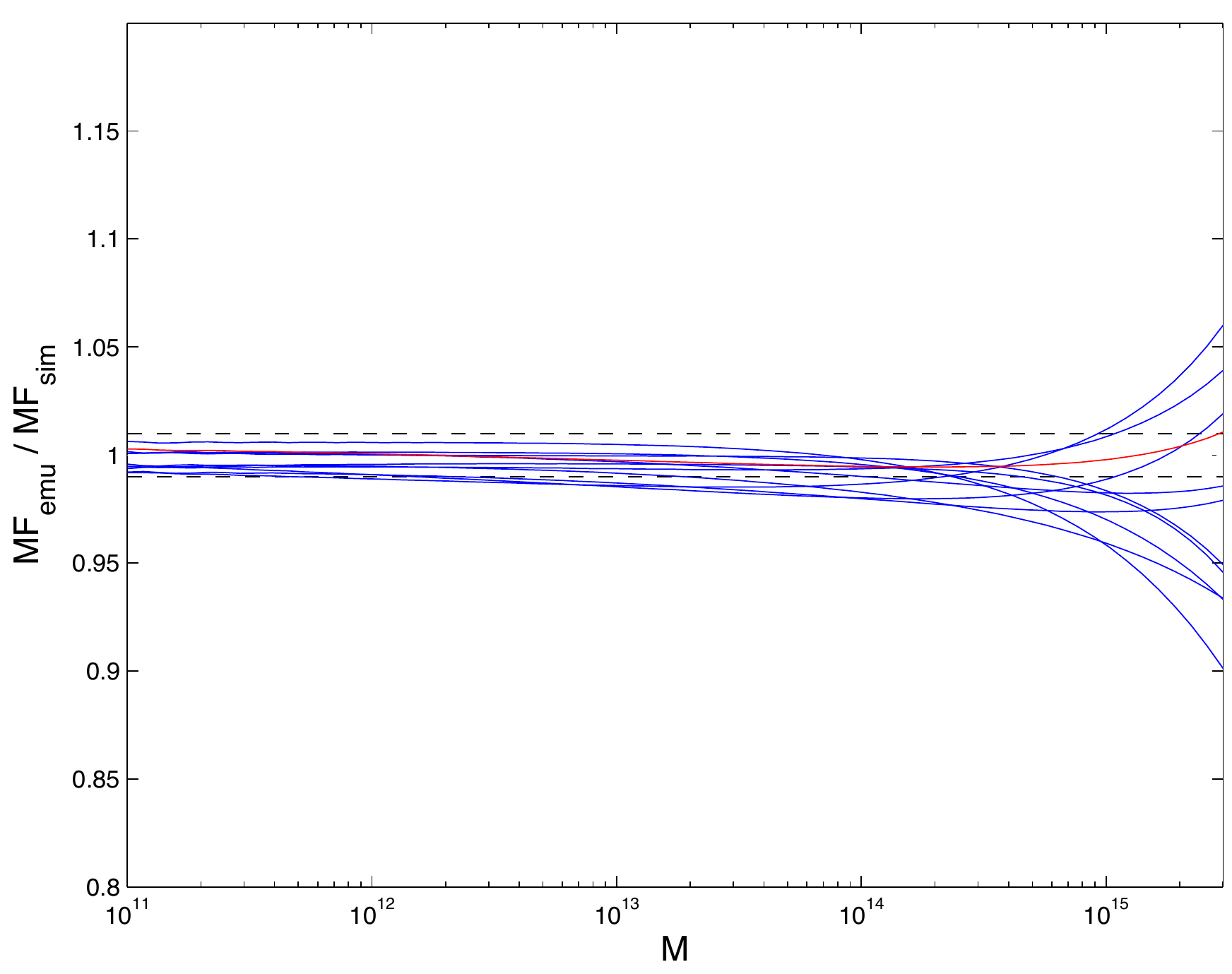} 
\includegraphics[width=1.7in]{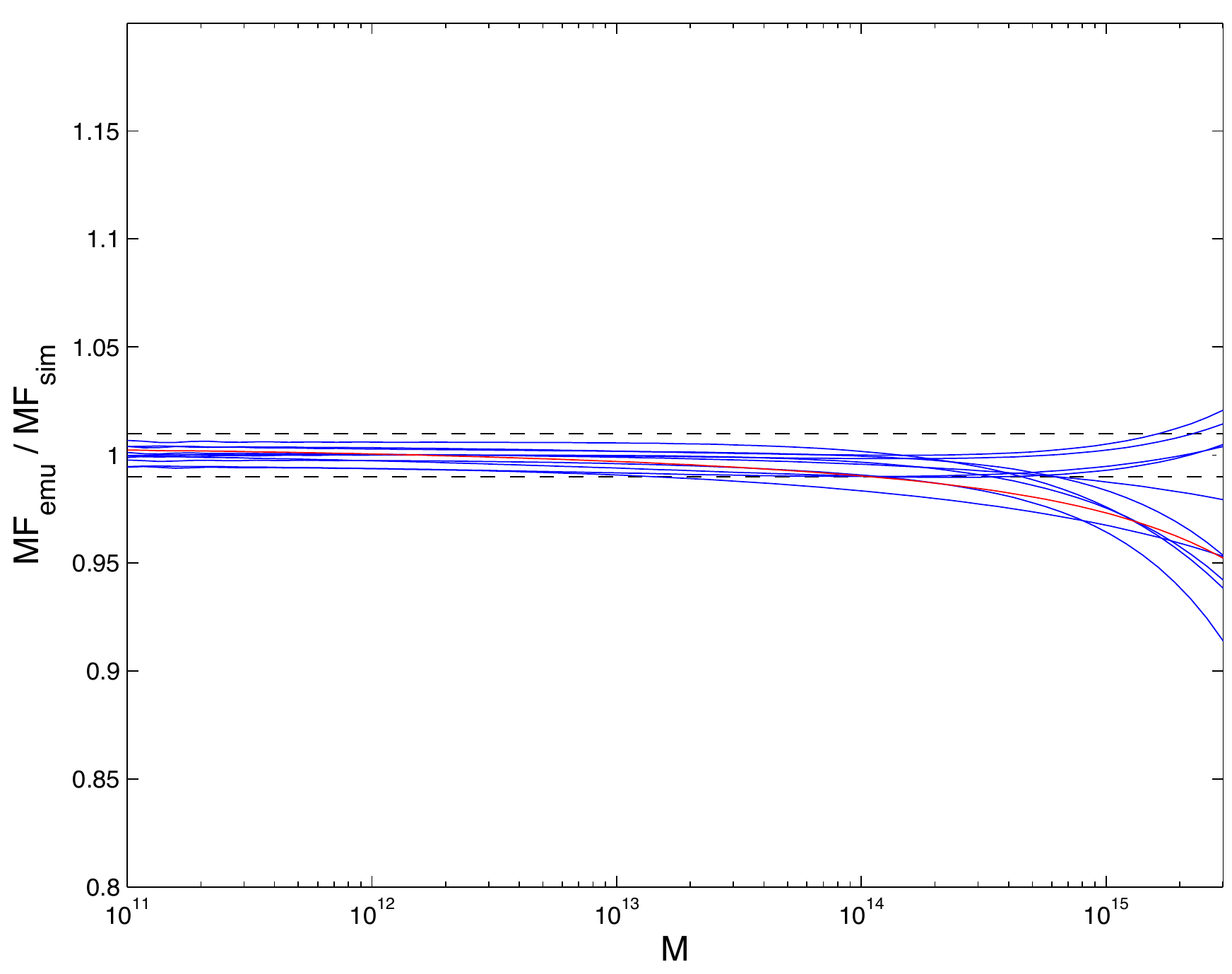}
\includegraphics[width=1.7in]{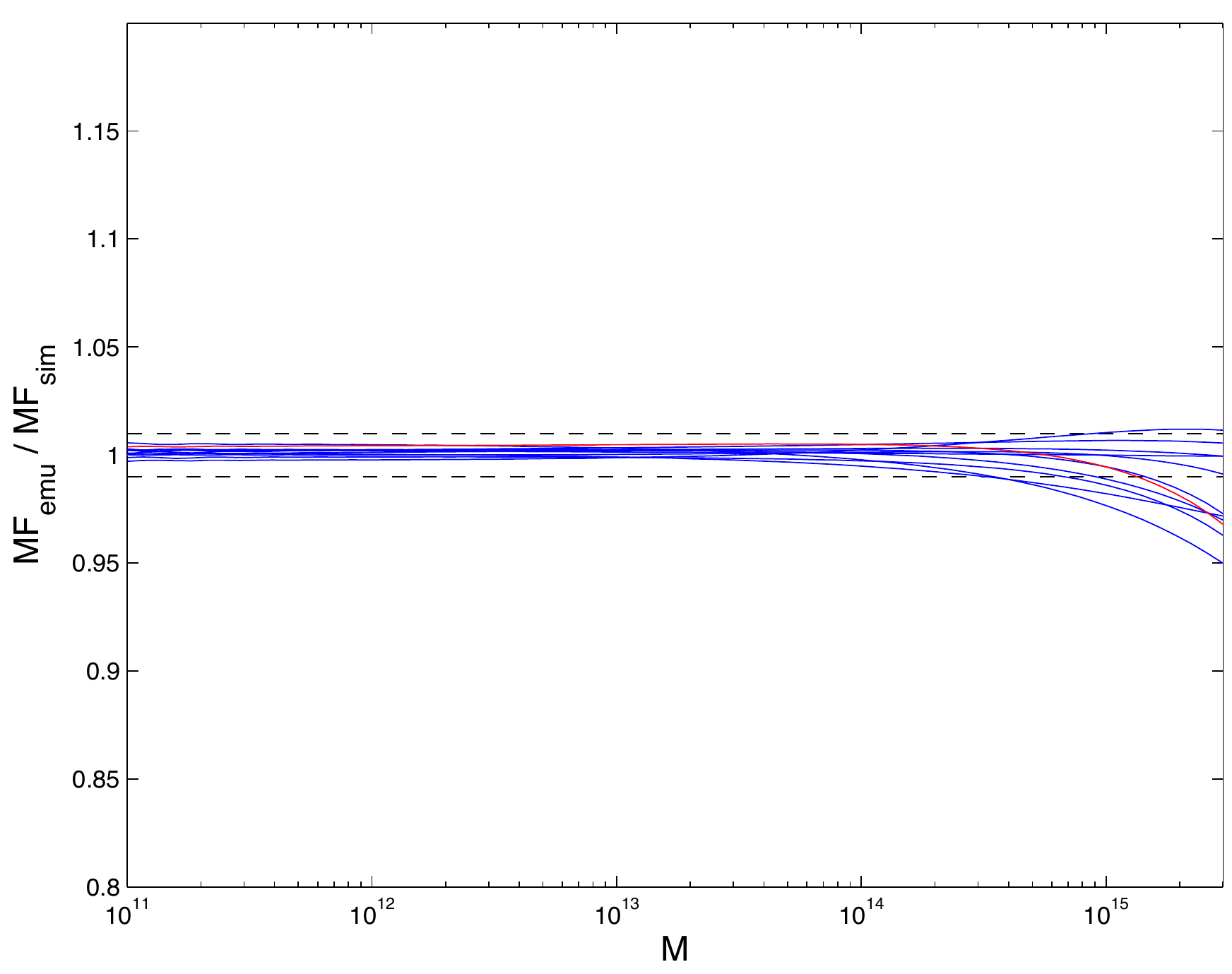}
\includegraphics[width=1.7in]{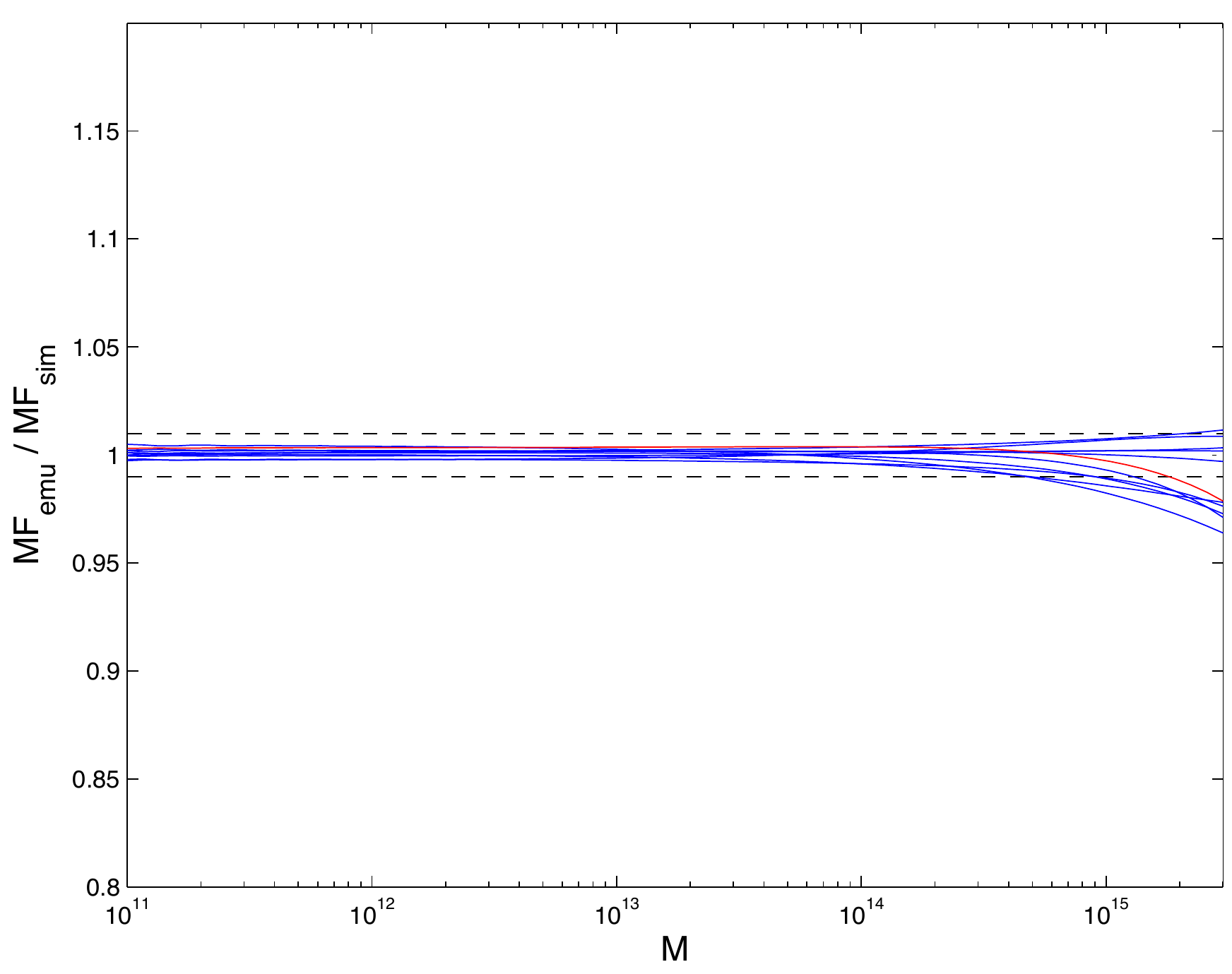}
\includegraphics[width=1.7in]{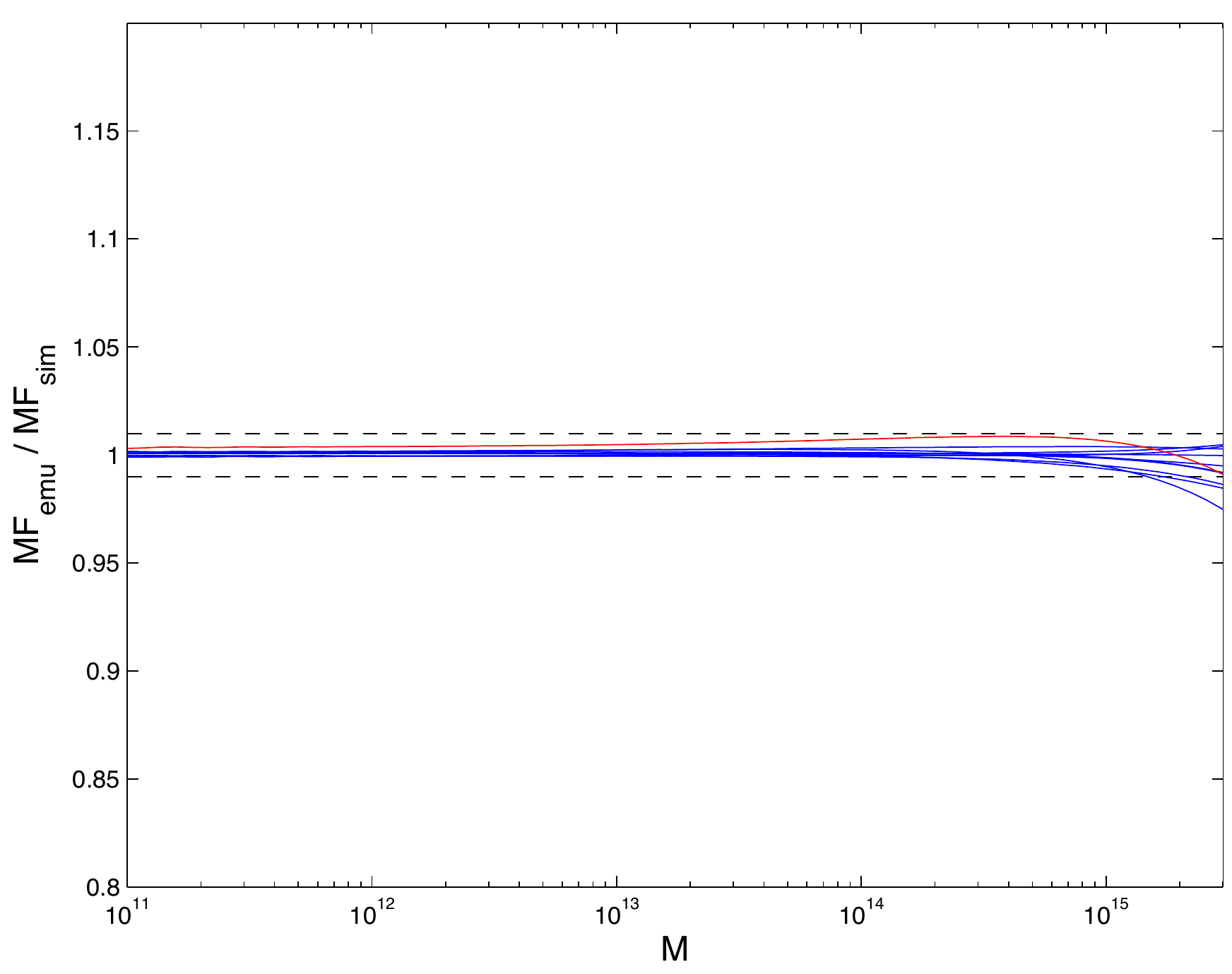}
\includegraphics[width=1.7in]{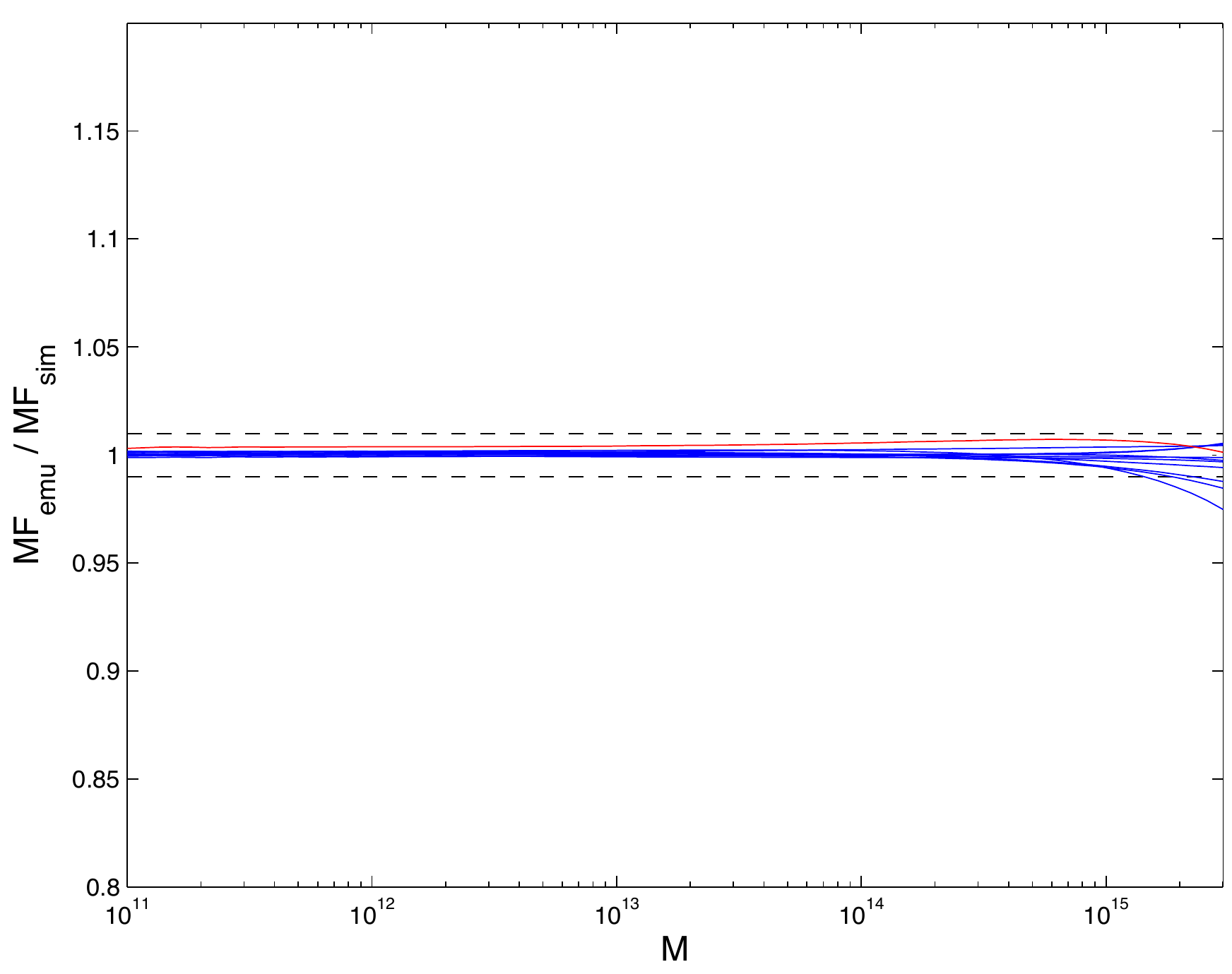}}
\caption{\label{emu_test_mf}Emulator test: An emulator for the mass
  function is built from up to 101 and 111 models, assuming
  universality. Next, the mass functions for 10 additional models are
  predicted with the emulator and compared to the prediction assuming
  universality. Shown is the ratio of prediction to ``truth'', the
  emulator predictions are accurate at the 1\% level (dashed line)
  over the full mass range once 101 (111) models are considered. The
  upper panels show the prediction for emulator based on 26 and 36
  models, the middle panels are based on 55 and 65 models, and for the
  predictions shown in the lower panels all 101 and 111 models were
  used. As before, the right panels include the addition 10 simulations with
  $m_\nu=0$.}
\end{figure}

Based on these mass function predictions, we build an emulator in the
same way as for the power spectrum.  Figure~\ref{emu_test_mf} shows
the results for the emulator predictions for the test models at
$z=0$. The first two panels show the results from an emulator based on
26 and 36 models. The accuracy of the emulator is already within 5\%,
a very encouraging result. Increasing the number of models to 55 (65
including the $m_\nu=0$ models) reduces the prediction error to $\sim
3\%$, and the final results for 101 and 111 models promise to yield
inaccuracies at the 1\% level over an 8-dimensional parameter
space. This leads us to the conclusion that if the mass function
prediction from the simulation can be provided at sufficient accuracy,
building highly accurate emulators will be straightforward.

\section{Beyond the Power Spectrum and Mass Function}
\label{beyond}

\subsection{Emulators}
In the previous sections we have shown examples of two precision
emulators to be built from the Mira-Titan Universe campaign targeted
to impacting analysis of data from ongoing and upcoming dark energy
surveys. The outlined simulation campaign lends itself to building
many more emulators. One such example is a concentration-mass ($c-M$)
emulator. In \cite{kwan13} it was shown that an accurate $c-M$
emulator can be built from a small set of simulations -- the paper was
based on an augmented set of Coyote Universe simulations.  The
Mira-Titan Universe volume and mass and force resolution is sufficient
to generate predictions for the $c-M$ relation over a mass range
covering groups to cluster scales.
\begin{figure}[t]
\centerline{
 \includegraphics[width=3.in]{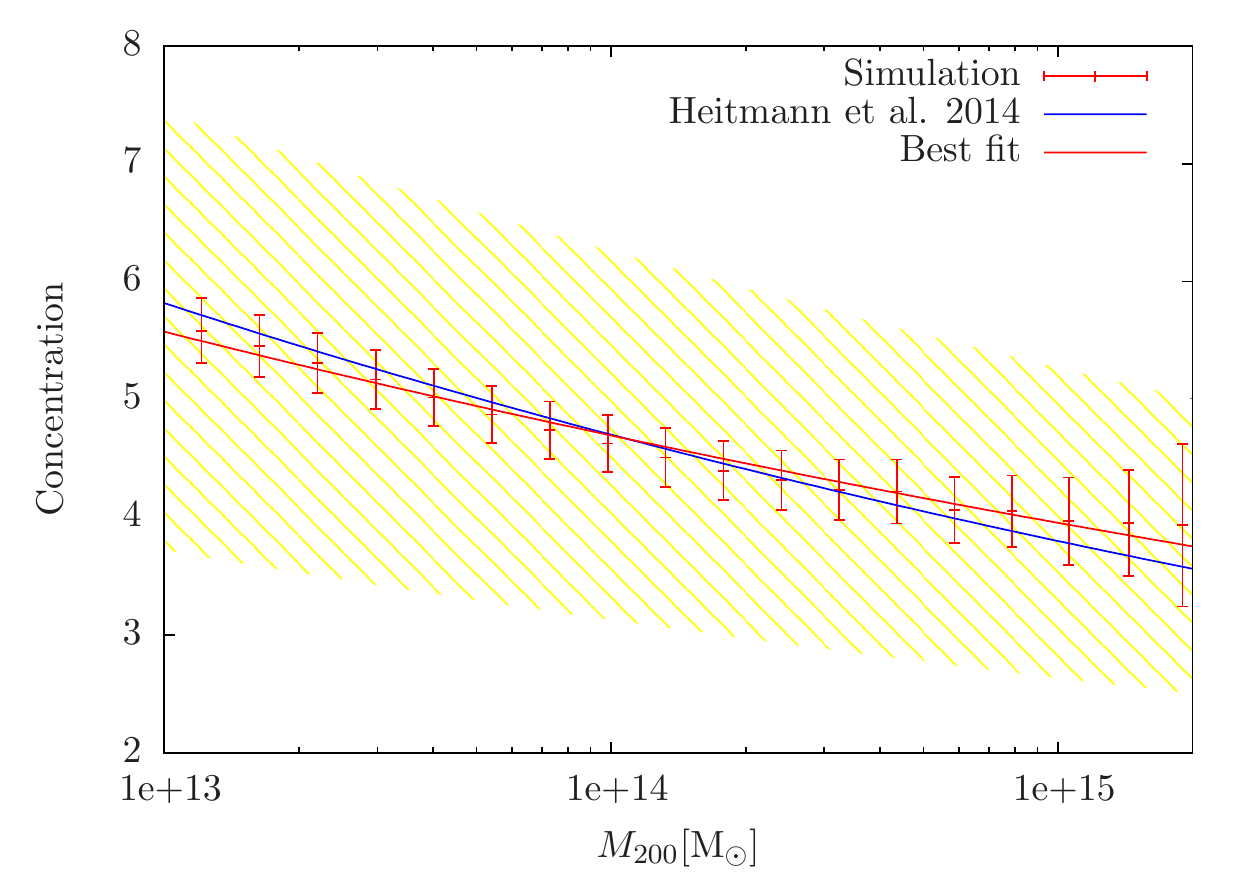}}
\caption{\label{cm_0}Concentration-mass relation as measured from the
  Mira-Titan Universe simulation (red points). The red line shows the
  best-fit power law through the simulation points. In blue we show
  the best fit $c-M$ relation found in the Q Continuum simulation
  (\citealt{QCont}), a simulation with almost 100 times better mass
  resolution, allowing the derivation of a fit out to much smaller
  halo masses.  The yellow shaded region shows the intrinsic
  scatter. Within this scatter, the predictions from the Mira-Titan
  Universe simulation agree very well with the high-mass resolution Q
  Continuum simulation.}
\end{figure}
Figure~\ref{cm_0} shows the $c-M$ relation measured from the
Mira-Titan Universe run at $z=0$. Given the intrinsic scatter in the
$c-M$ relation, the statistics obtained from the Mira-Titan Universe
simulation are more than sufficient to derive accurate $c-M$ relation
predictions over the mass range shown. Each halo here is resolved with
at least 1000 particles to provide well-sampled halos when determining
the concentration (for details about the concentration measurements, 
see \citealt{bhattacharya13}). For comparison, we show the best fit
power law derived from the Q Continuum simulation~\cite{QCont}, a
simulation with much higher mass resolution. The agreement is very
good.

Another example is an emulator constructed for the galaxy power
spectrum. In \cite{kwan14}, the Mira-Titan Universe simulation was
used to build a galaxy power spectrum emulator covering a wide range
of HOD models. With the complete suite of models we will be able to
extend that work to not only cover different HOD models but to also
provide predictions across cosmologies.  Extending our work into
redshift space is another obvious direction, as redshift space
distortion power spectrum and correlation function emulators can be
easily obtained based on the simulations outlined here.

\subsection{Sky Maps}
\label{maps}
Beyond emulators, the simulation suite described here will also be
very valuable for producing synthetic sky maps. For the LSST Dark
Energy Science Collaboration we have generated an HOD-based galaxy
catalog for the large scale structure working group. The size and
resolution of the main simulations will also be sufficient to build
weak lensing maps for surveys such as DES. Maps tracing the kinematic
Sunyaev-Zel'dovich effect are currently being developed. We are
planning to publicly release these maps to the community.

\section{Conclusion}
\label{conclusion}

In this paper we have introduced the Mira-Titan Universe, a new
simulation campaign to provide accurate predictions for a range of
cosmological observables and for generating sky maps in different
wavebands.  This work is a major new extension of our previous Coyote
Universe project in several key respects.
\begin{enumerate}
\item The cosmological parameter space covered is extended by three
  parameters, $w_a$, $h$, and $m_\nu$, all of which will play a
  central role in future analyses of dark energy surveys. We have
  shown that despite the extended parameter space the number of models
  needed to build reliable emulators is still manageable.
 
\item A new design strategy has been implemented using space-filling
  lattices. This strategy allows us to improve the emulator quality by
  systematically adding more simulations to an existing latice design.
  Our new strategy has the benefit of explicitly enforcing
  space-filling properties in intermediate stages as well as in the
  final design.  We would like to stress that this procedure is highly
  nontrivial -- extensive tests showed that simply adding independent
  space-filling designs (as in conventional Latin hypercube sampling)
  will not lead to the desired results.

\item We have extended our investigations beyond the power spectrum
  and demonstrated that our design will also provide high-accuracy
  emulators for the mass function.  In addition, the simulation set
  will allow us to build a range of other emulators, for the halo
  concentration-mass relation, galaxy correlation functions and power
  spectra, and redshift space distortions.

\item We have demonstrated with a set of high resolution simulations
  that the needed requirements for such a simulation suite can be
  met. These simulations can then be also used for a range of other
  investigations, in particular building mock catalogs for different
  wavebands.
\end{enumerate}

\begin{acknowledgments}

  The authors would like to thank Juliana Kwan, Amol Updahye, and
  Martin White for many useful conversations.

  Part of this research was supported by the DOE under contract
  W-7405-ENG-36.  Argonne National Laboratory's work was supported
  under the U.S. Department of Energy contract
  DE-AC02-06CH11357. Partial support for HACC development was provided
  by the Scientific Discovery through Advanced Computing (SciDAC)
  program funded by the U.S. Department of Energy, Office of Science,
  jointly by Advanced Scientific Computing Research and High Energy
  Physics.  KH was supported in part by NASA. RB acknowledges partial
  support from the Washington Research Foundation Fund for Innovation
  in Data-Intensive Discovery and the Moore/Sloan Data Science
  Environments Project at the University of Washington.
  
  This research used resources of the ALCF, which is supported by
  DOE/SC under contract DE-AC02-06CH11357 and resources of the
  National Energy Research Scientific Computing Center, a DOE Office
  of Science User Facility supported by the Office of Science of the
  U.S. Department of Energy under Contract No. DE-AC02-05CH11231.
\end{acknowledgments}

\appendix
\label{appendixa}

In this Appendix we list the test cosmologies that were used to verify
the accuracy of the emulators discussed in the paper.

\begin{table*}
\caption{Validation Set\label{tab2}}
\begin{center}
\begin{tabular}{ccccccccc}
Model & $\omega_m$ & $\omega_b$ & $\sigma_8$ &   $h$ & $n_s$    & $w_0$ & $w_a$  &$\omega_\nu$ \\
\hline\hline
E001 &0.1433 &  0.02228 &  0.8389 &  0.7822 &  0.9667 &  -0.8000 &  -0.0111 & 0.008078 \\
E002 &0.1333 &  0.02170 &  0.8233 &  0.7444 &  0.9778 &  -1.1560 &-1.1220 & 0.005311 \\
E003 &0.1450 &  0.02184 &  0.8078 &  0.6689 &  0.9000 &  -0.9333 &  -0.5667 &  0.003467 \\
E004 &0.1367 &  0.02271 &  0.8544 &  0.8200 &  0.9444 &  -0.8889 &  -1.4000 & 0.002544 \\
E005 &0.1400 &  0.02257 &  0.7300 &  0.7067 &  0.9889 &  -0.9778 &  -0.8444 &  0.009000 \\
E006 &0.1350 &  0.02213 &  0.8700 &  0.7633 &  0.9111 &  -1.0220 & 0.5444 &  0.000700 \\
E007 &0.1383 &  0.02199 &  0.7456 &  0.6500 &  0.9556 &  -1.1110 & 1.1000 & 0.007156 \\
E008 &0.1300 &  0.02286 &  0.7922 &  0.8011 &  1.0000 & -1.0670 & 0.2667 &  0.006233 \\
E009 &0.1417 &  0.02300 &  0.7767 &  0.7256 &  0.9222 &  -0.8444 &  0.8222 &  0.004389 \\
E010 &0.1317 &  0.02242 &  0.7611 &  0.6878 &  0.9333 &  -1.2000 &-0.2889 &  0.001622
\end{tabular}
\end{center}
\end{table*}

\end{document}